\documentclass[journal]{IEEEtran}
\usepackage{cite}
\usepackage{amsmath,amssymb,amsfonts}

\usepackage{algorithm} 

\usepackage{subfigure}

\usepackage{textcomp}
\usepackage{bm}
\usepackage{amsmath,amsfonts}
\usepackage{algorithmic}
\usepackage{array}
\usepackage[caption=false,font=normalsize,labelfont=sf,textfont=sf]{subfig}
\usepackage{textcomp}
\usepackage{stfloats}
\usepackage{url}
\usepackage{verbatim}
\usepackage{graphicx}
\usepackage{color}
\usepackage{siunitx}
\usepackage{mhchem}
\usepackage{booktabs}
\usepackage{tabularx}

\def\BibTeX{{\rm B\kern-.05em{\sc i\kern-.025em b}\kern-.08em
    T\kern-.1667em\lower.7ex\hbox{E}\kern-.125emX}}
\usepackage{balance}  
\begin{document}
\title{Trajectory Design and Resource Allocation for Multi-UAV-Assisted Sensing, Communication, and Edge Computing Integration} 
\author{Sicong Peng,
Bin Li,~\IEEEmembership{Member,~IEEE},
Lei Liu,~\IEEEmembership{Member,~IEEE},
Zesong Fei,~\IEEEmembership{Senior Member,~IEEE}, \\
and Dusit Niyato,~\IEEEmembership{Fellow,~IEEE}

\thanks{Sicong Peng and Bin Li are with the School of Computer Science, Nanjing University of Information Science and Technology, Nanjing 210044, China (e-mail: sicong.peng@nuist.edu.cn; bin.li@nuist.edu.cn).}
\thanks{Lei Liu is with the Guangzhou Institute of Technology, Xidian University, Guangzhou 510555, China (e-mail: tianjiaoliulei@163.com).}
\thanks{Zesong Fei is with the School of Information and Electronics, Beijing Institute of Technology, Beijing 100081, China (e-mail: feizesong@bit.edu.cn).}
\thanks{Dusit Niyato is with the College of Computing and Data Science, Nanyang Technological University, Singapore (e-mail: dniyato@ntu.edu.sg).}
}

\maketitle
\begin{abstract}
In this paper, we propose a multi-unmanned aerial vehicle (UAV)-assisted integrated sensing, communication, and computation network. Specifically, the treble-functional UAVs are capable of offering communication and edge computing services to mobile users (MUs) in proximity, alongside their target sensing capabilities by using multi-input multi-output arrays. For the purpose of enhance the computation efficiency, we consider task compression, where each MU can partially compress their offloaded data prior to transmission to trim its size. The objective is to minimize the weighted energy consumption by jointly optimizing the transmit beamforming, the UAVs' trajectories, the compression and offloading partition, the computation resource allocation, while fulfilling the causal-effect correlation between communication and computation as well as adhering to the constraints on sensing quality. To tackle it, we first reformulate the original problem as a multi-agent Markov decision process (MDP), which involves heterogeneous agents to decompose the large state spaces and action spaces of MDP. Then, we propose a multi-agent proximal policy optimization algorithm with attention mechanism to handle the decision-making problem. Simulation results validate the significant effectiveness of the proposed method in reducing energy consumption. Moreover, it demonstrates superior performance compared to the baselines in relation to resource utilization and convergence speed.
\end{abstract}      


\section{Introduction}
\IEEEPARstart{W}{ith} the incorporation of communication and information technologies, the nodes in future 6G networks are expected to transcend their traditional communication roles and seamlessly integrate multiple functionalities \cite{9606720}. To meet these demands, it is essential to significantly improve the quality of service and application experience in the 6G networks \cite{9328305}. However, due to the coexistence of communication, sensing, and computation functionalities, achieving such an evolution of 6G wireless networks from ``connected things" to ``connected intelligence" is non-trivial. For instance, the newly increased demand for radio sensing will place an even greater burden on already scarce spectrum resources, and the additional requirements for rapid offloading will soon encounter performance bottlenecks in congested core networks.

In light of the thorny problems encountered above, it naturally leads to the idea that the future network nodes are poised to transcend their conventional role of providing communication services and progress towards a cohesive framework that integrates communication, sensing, and computation (ISCC) functions \cite{9282063}. Benefiting from the advancements in network nodes and computational technologies, the integration of mobile edge computing (MEC) with integrated communication and sensing technology has arisen as a promising approach \cite{9520318}. Specifically, MEC has emerged as a prevalent paradigm that opens new possibilities for facilitating low-latency, computation-intensive applications to be executed on the infrastructure side. Concurrently, due to the shared hardware infrastructures and channels with wireless communication, wireless radio sensing, leveraging wireless infrastructures to collect environmental characteristices, has arisen as a highly promising sensing solution, which can achieve mutual benefit with communication \cite{9737357}.

Generally, the performance of ISCC can be significantly impacted by blockage and inadequate coverage in areas located far from the servers, such as disaster-stricken regions, remote locations, and crowded areas \cite{9522072}. To address this dilemma, unmanned aerial vehicles (UAVs) have been established as a bright technology for assisting wireless communications by leveraging their enhanced mobility, versatile deployment, and cost-effectiveness \cite{8579209},\cite{9943536}. By cooperatively carrying MEC servers and establishing strong line-of-sight connections, UAVs can fly  in close proximity to MUs to offload part of the tasks and play a crucial role in guaranteeing the reliable connectivity of MUs, which is an expedient and cost-efficient deployment method for computation offloading with minimal delivery latency and demanding high data transmission rate \cite{9672750}. The inclusion of UAVs further enhances the dynamic nature of ISCC networks, thus prompting the utilization of online optimization techniques in MEC services. In particular, the fusion of artificial intelligence methodologies provides superior performance compared to traditional offline optimization approaches, offering faster decision-making and greater adaptability to the dynamic environment.

Meanwhile, data compression (DC) has become pervasive and finds applications across various domains \cite{data}, by which, we can save storage space, reduce transmission latency, and improve transmission efficiency. On the other hand, MIMO technology has been widely utilized in ISAC systems to enhance sensing performance while achieving simultaneous gains in multi-MU communications \cite{9648018}. When these two technologies are applied to UAV-assisted ISCC systems, high-mobility UAVs under DC and MIMO will provide superior quality of service.

\subsection{Related Work}
\subsubsection{Researches into MEC}Owing to the advantageous characteristics of MEC, there has been extensive discourse on maximizing energy efficiency, minimizing transmission latency, and maximizing computation rate. Several recent studies have explored video compression offloading in MEC systems to minimize overall latency \cite{8387798}, while others have investigated energy consumption minimization through DC in MEC networks \cite{8635566}. However, the aforementioned works primarily consider fixed base station (BS) situations. To harness the capabilities of UAVs in diverse scenarios, the collaborative deployment and resource allocation in UAV-assisted disaster emergency communication were the focus in \cite{9453853}, which was addressed by iterative approaches. Furthermore, multi-agent deep reinforcement learning (MADRL), a powerful technique for training efficient decentralized models, attracts significant attention in various aspects. More specifically, in \cite{9622148}, the authors focused on minimizing latency of execution in a heterogeneous edge-cloud network that incorporates both BSs and UAVs via the MADRL approach. The work of \cite{9451579} concentrated on the configuration of resource allocation and task offloading among a mirrored MEC system, utilizing an MADRL algorithm. 

\subsubsection{Researches into ISCC} In the literature of ISCC networks, some efforts have put emphasis on joint resource scheduling to boost the performance of sensing, communication and computation. For instance, in \cite{9828481}, a multi-objective framework was proposed to optimize beamforming design from the aspect of maximizing overall performance and minimizing transmit power consumption. Building upon this, a multi-objective problem was designed in \cite{9729765} involving radar beampattern and computational energy minimization in ISCC systems. For the application of UAV in ISCC, apart from designing system resource allocation and beamforming, a comprehensive investigation into optimizing UAV trajectory is required. In \cite{9916163}, a framework was proposed to coordinate UAV trajectories and beamforming strategies in anticipation of improving the property of integration of sensing and communication network. The authors of \cite{9963915} studied the joint optimization problem of UAV trajectory, resource allocation and transmit power to strike the trade-off between radar sensing and communication performance for all UAVs. Designed to enhance communication and sensing coverage to a greater extent, the authors in \cite{9293257} investigated a joint optimization among UAV flight trajectories, MU association and resource allocation to maximize the integrated gains of communication and sensing. In \cite{9739676}, the adaptability in sensing duration was proposed to mitigate wastage of UAV-assisted ISCC's resource and excessive sensing. The authors in \cite{9656117} investigated the cooperative behavior exhibited by UAVs and intelligent reflecting surface with the objective of maximizing the average secrecy rate and enhancing the overall service performance. The authors in \cite{9729746} designed a sensing-control threshold, which utilizes the closed-form relationship of sensing-control pattern and communication rate, to guarantee the communication rate requirement and meanwhile maintain satisfactory motion performance of UAV. To further reduce the propulsion energy consumption of UAVs, a UAV trajectory planning problem under a novel BS-UAV bistatic radar platform was proposed in \cite{9847217}. In addition, the exploration of intelligent methods in the realm of ISCC is still at a nascent stage, offering potential avenues for further research and development.

\subsection{Contributions and Organization}
To fully explore the potential of UAV-aided ISCC networks and realize flexible services, this paper first introduces data compression into ISCC networks, aiming to significantly reduce resource consumption. Moreover, we fully capitalize on UAV's high mobility, effectively utilizing it as an aerial edge server to deliver communication services and computing resources to MUs. In addition, this paper adopts a newly proposed multi-agent proximal policy optimization (MAPPO) algorithm to solve the joint optimization problem of heterogeneous network resources and real-time policies. The main contributions of this work are summarized as follows:

\begin{enumerate}
    \item  We investigate a multi-UAV-assisted ISCC network that leverages UAV and DC technique to enable the integration of multiple functionalities. Taking into account the energy limitations of both MUs and UAVs, we establish a formulation for the problem of minimizing weighted energy consumption by collaboratively designing MU association, compression proportion, offloading proportion, multi-input multi-output (MIMO) beamforming, resource allocation, and trajectory control on sensing, computation and communication.
    \item  To tackle the high complexity of the real-time problem, we opt to reframe it as a Markov decision process (MDP) and utilize the MADRL framework, which accommodates heterogeneous agents to effectively handle the challenges posed by the high-dimensional state and action spaces. To promote collaborative decision-making, we propose to apply the MAPPO algorithm through a distributed and online approach.
    \item  To enhance the training performance and expedite convergence speed, we incorporate attention mechanism and Beta distribution into the critic network and actor network, respectively. Via simulation results, the fast convergence in training and the effectiveness of our proposed scheme in optimizing the problem have been verified.
\end{enumerate}

The remainder of this paper is structured as follows. In Section \ref{s:sys}, we elaborate on the system model of the UAV-assisted ISCC network and formulate the objective problem. The design of the ATB-MAPPO algorithm is presented in Section \ref{s:proposed}. Subsequently, Section \ref{s:simulation} presents the simulation results, and we provide the concluding remarks in Section \ref{s:conclusion}.

\textit{Notations:} In this paper, matrices and vectors are denoted by boldface capital boldface and lower case letters, respectively. $\left\lvert \cdot \right\rvert $ and $\left\lVert \cdot \right\rVert $ denote the absolute value of a complex scalar and the Euclidean norm of a vector, respectively. $\text{h}^T$ and $\text{h}^H$ represent the transpose and conjugate transpose of vector $\text{h}$. $\mathbb{C} ^{M\times N}$ and $\mathbb{R} ^{M\times N}$ denote the set of $M \times N$ complex-valued and real-valued matrices, respectively.

\begin{figure}[t]
    \centering
    \includegraphics[width=\columnwidth]{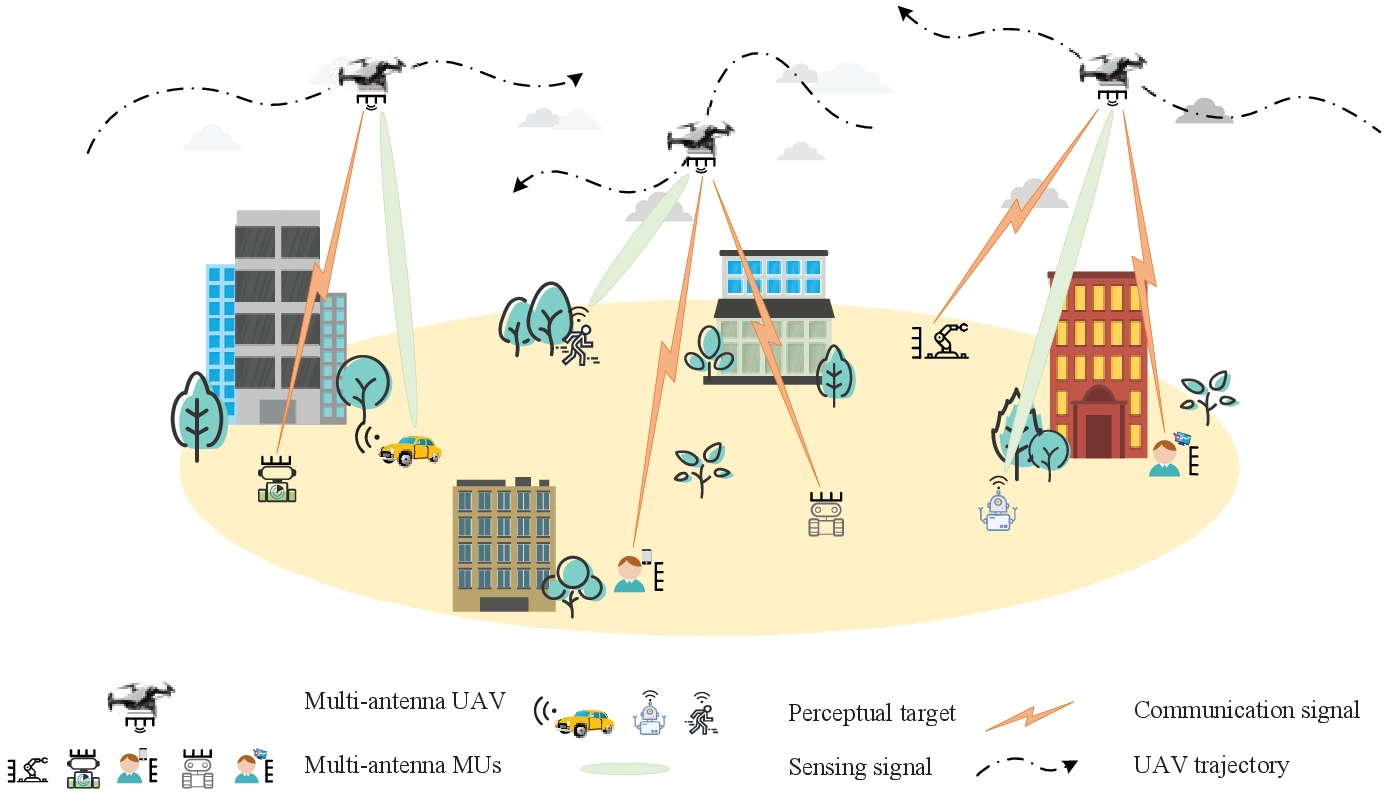}
    \caption{The system model of multi-UAV-assisted ISCC network.}
    \label{fig:MIMO-UAVs-ISCC}
\end{figure}

\section{System Model and Problem Formulation}\label{s:sys}
As shown in Fig. \ref{fig:MIMO-UAVs-ISCC}, we consider a multi-UAV-assisted ISCC network containing $K$ MUs with $W_T$ antennas and $M$ UAVs with $W_R$ antennas, which are defined as the set $\forall k \in \mathcal{K}\triangleq\{1,2,\cdots,K\}$ and the set $\forall m \in \mathcal{M}\triangleq\{1,2,\cdots,M\}$, respectively. The UAV is furnished with triple function systems, featuring radar detection, wireless communication as well as edge computing capabilities. The UAV's radar is capable of sensing the environment as well as communicating with MUs to swap control data and fundamental status revisions. Additionally, the UAVs deploy MEC servers to efficiently process computationally-intensive tasks generated by MUs. Simultaneously, we express the flight period of UAVs as $T=\delta_{t}T_{\text{ts}}$, where we denote the set of time slots as $\forall t \in \mathcal{T}\triangleq\{1,2,\cdots,T_{\text{ts}}\}$ and $\delta_{t}$ is expressed as the length of time slot.

The three-dimensional coordinate system is adpoted to label the position of UAVs and MUs. To be more specific, the coordinates of UAV $m$ can be expressed as ${\bf q}_{m}[t]=[x_{m}[t],y_{m}[t],H]^T$, where $H$ is the flying height, and that of MU $k$ is defined as ${\bf w}_{k}[t]=[x_{k}[t],y_{k}[t]]^T$. Since the limited computation resources of MUs, their tasks can be offloaded to the associated UAVs. Therefore, the association factor between UAV $m$ and MU $k$ is defined as follows:

\begin{equation}
    {\sum_{m=1}^M \alpha_{k,m}[t]\leq1,\sum_{k=1}^K \alpha_{k,m}[t]\leq1,  \forall m \in \mathcal{M}, k \in \mathcal{K} ,}\label{con:1}
\end{equation}
\begin{equation}
    {\alpha _{k,m}[t]\in\{0,1\}.\label{con:2}}
\end{equation}

When $\alpha _{k,m}[t]=1$, the state indicates a successful association between MU $k$ and UAV $m$, while $\alpha _{k,m}[t]=0$ represents an unpaired state. Note that all MUs are initially located randomly, and navigate following the Gaussian-Markov random model\cite{10086052}. Specifically, by assuming that the locations of all MUs remain static during a time slot, the direction $\theta_{k}[t]$ and the velocity $v_k[t]$ of MU $k$ are given by
\begin{equation}
    \theta_k[t]=\mu_{2}\theta_k[t-1]+(1-\mu_{1})\bar{\theta}+\sqrt{1-\mu_{2}^{2}}\Psi_{k}, \label{5}
\end{equation}
\begin{equation}
    v_k[t]=\mu_{1}v_k[t-1]+(1-\mu_{1})\bar{v}+\sqrt{1-\mu_{1}^{2}}\Phi_{k}, \label{4}
\end{equation}
where $0\leq \mu_{1},\mu_{2}\leq1$ are utilized to adjust the influence of the prior state, $\bar{\theta}$ and $\bar{v}$ denote the mean direction and velocity of MUs, respectively. Additionally, $\Phi_{k}$ and $\Psi_{k}$ are generated from two separate Gaussian distributions. The former has a mean $\bar{\xi}_{v_k}$ and variance $\zeta_{v_k}^{2}$, while the mean-variance combination of the latter is $(\bar{\xi}_{\theta_k},\zeta_{\theta_k}^{2})$. These mean-variance combinations determine the characteristics of the generated values for each variable. According to \eqref{4} and \eqref{5}, the location of MU $k$ can be updated as
\begin{equation}
    x_{k}[t]=x_{k}[t-1]+v_k[t-1]\cos(\theta_k[t-1])\delta_{t}, 
\end{equation}
\begin{equation}
    y_{k}[t]=y_{k}[t-1]+v_k[t-1]\sin(\theta_k[t-1])\delta_{t}.
\end{equation}

\subsection{Channel Model}
In practical scenarios, the channels between UAVs and MUs are affected by various obstacles such as buildings and trees, which result in blockages and the presence of numerous scattering components. As such, the channel model between UAV and MU is typically characterized by a Rician fading channel model, where the representation of the channel model between MU $k$ and UAV $m$ is

\begin{equation}
    \mathbf{H}_{k,m}[t]\!=\!\!\!\sqrt{\frac{\rho}{d_{k,m}^{2}[t]}}\!\!\left(\!\!\sqrt\frac{\epsilon}{\epsilon+1}\!\bar{\mathbf{H}}_{k,m}[t]\!+\!\sqrt\frac{1}{\epsilon+1}\!\tilde{\mathbf{H}}_{k,m}[t]\!\!\right). \label{7}
\end{equation}
In \eqref{7}, $d_{k,m}^{2}[t]={\left\lVert {\bf q}_{m}[t]-{\bf w}_{k}[t]\right\rVert}^2+H^2$, $\rho$ represents the channel power gain at the reference distance, and $\epsilon$ is the Rician factor specifying the power ratio. Additionally, $\bar{\mathbf{H}}_{k,m}[t]\in\mathbb{C}^{W_R\times W_T}$ and $\tilde{\mathbf{H}}_{k,m}[t]\in\mathbb{C}^{W_R\times W_T}$ are equivalent to the line-of-sight channel and the non-line-of-sight one and $\tilde{\mathbf{H}}_{k,m}[t]\sim \mathcal{CN}\left(0,\mathbf{I}_{W_R} \right)$.

To accomplish the signal transmission between MUs and UAVs, we denote the transmit signal of computation MU $k$ as $\bar{{\bf s}}_{k,m}[t]=\sqrt{P_{k}[t]}{\bf s}_{k,m}[t]$ for delivering the information stream ${\bf s}_{k,m}[t]$ to UAV $m$ with transmit power $P_{k}[t]$. Subsequently, the UAV $m$ receives the superimposed computation offloading signals ${\bf y}_{m,c}[t]$ from the MUs, which can be formulated as 
\begin{equation}
    {\bf y}_{m,c}[t]=\sum_{k = 1}^{K}\alpha_{k,m}[t]\sqrt{P_{k}[t]}\mathbf{H}_{k,m}[t]{\bf s}_{k,m}[t].
\end{equation}

Then, we analyze the radar sensing signal, assuming that the prior information of target sources is gathered by UAVs. Specifically, the information of the target is acquired through estimation results from the previous sensing phase\cite{9996408}. This prior knowledge enables the UAV's radar to distinguish between target and other UAVs' clutter sources, thereby facilitating more accurate target detection and localization. Moreover, we suppose that the Doppler shift induced by moving targets is constant over the duration of the radar pulse repetition so that the range-Doppler parameters are fully compensated. Denote ${\bf x}_{m}[t]=\mathbf{W}_{m}[t]{s}_{m}[t]$ as the transmit signal at UAV $m$ to sense the target, where $\mathbf{W}_{m}[t]\in\mathbb{C}^{W_{R}\times 1}$ represents the radar sensing beamformer, and ${s}_{m}[t]$ is a radar waveform. Therefore, the echo received by UAV $m$ can be given by
\begin{equation}
    {\bf y}_{m,r}[t]=\varphi_{m}\mathbf{A}_{m}[\xi_{m}]{\bf x}_{m}[t]+\sum_{i=1,i\neq m}^{M}\Theta_{m,i} {\bf x}_{i}[t]+ {\bf z}_{m}[t]. 
\end{equation}
Denoting $\mathbf{A}_{m}[\xi_{m}]={\bf a}_{R,m}(\xi_{m}){\bf a}_{T,m}^{\text H}(\xi_{m})$, in which ${\bf a}_{T,m}(\xi_{m})\in \mathbb{C}^{W_{R}\times 1}$ is the transmit array steering vector of the radar for UAV $m$ and ${\bf a}_{R,m}(\xi_{m})\in \mathbb{C}^{W_{R}\times 1}$ represents the receive one. In addition, $\xi_{m}$ is a direction from target to radar, $\Theta_{m,i}$ is the channel interference between UAVs, $\varphi_{m}$ represents the Doppler frequency shift, and ${\bf z}_{m}[t]\sim \mathcal{CN}\left(0,\sigma_{2}\mathbf{I}_{W_R} \right)$ represents an additive white Gaussian noise. It is noteworthy that, for both planar wave and uniform linear array deployed on UAVs, the steering vector can be expressed as follows: 
\begin{equation}
    \begin{aligned}
        {\bf a}_{m}(\xi_{m})&={\bf a}_{T,m}(\xi_{m})={\bf a}_{R,m}(\xi_{m}) \\
        &={\left[1,e^{j\frac{2\pi }{\lambda }d_{m}\sin(\xi_{m})},\cdots ,e^{j\frac{2\pi }{\lambda }d_{m}\sin(\xi_{m})(W_{R}-1)}\right]}^{\text T},
    \end{aligned} 
\end{equation}
where $d_{m}=\lambda/2$ is the wavelength of the signal, and $\lambda$ is the UAVs' antenna spacing. 

\subsection{Communication Model}
To achieve efficient computation task offloading, the achievable communication rate plays a crucial role in determining system performance. For decoding the computation offloading signal transmitted by MUs at UAV $m$, we exploit successive interference cancellation, which permits the receiver to decode individual stream in a sequential manner, subsequently nullifying the impact of the already decoded streams from the received signal \cite{9424021}. Denoting ${\bf W}_{k,m}[t]\in\mathbb{C}^{W_{R}\times 1}$ as the receive beamforming vector between UAV $m$ and MU $k$. The signal received by UAV $m$ from MU $k$ is written as
\begin{equation}
    \begin{aligned}
        {\bf x}_{k,m}[t] &=\alpha_{k,m}[t]\sqrt{P_{k}[t]}{\bf W}_{k,m}^{\text H}[t]\mathbf{H}_{k,m}[t]{\bf s}_{k,m}[t]\\ &+\sum_{i=1,i\neq k}^{K}\sum_{j=1}^{M}\alpha_{i,j}[t]\sqrt{P_{i}[t]}{\bf W}_{k,m}^{\text H}[t]\mathbf{H}_{i,j}[t]{\bf s}_{i,j}[n] \\
        &+{\bf W}_{k,m}^{\text H}[t](\varphi_{m}\mathbf{A}_{m}[\xi_{m}]+\!\!\!\!\!\sum_{i=1,i\neq m}^{M}\!\!\varphi_{i}\mathbf{A}_{i}[\xi_{i}])\mathbf{W}_{m}[t]{\bf s}_{m}[t]\\ &+{\bf W}_{k,m}^{\text H}[t]{\bf z}_{k}[t]. \label{11}
    \end{aligned}
\end{equation}

For notational convenience, $\varphi_{m}\mathbf{A}_{m}[\xi_{m}]$ and $\sum_{i=1,i\neq m}^{M}\Theta_{m,i}$ in the above formula are substituted with ${\bf \Xi}_{m}$ and ${\bf \Xi}_{i}$, respectively. Accordingly, the signal-to-interference-plus-noise ratio (SINR) is calculated as
\begin{equation}
    \mathbf{\Upsilon}_{k,m}[t]=P_{k}[n]{\bf W}_{k,m}^{\text H}[t]\mathbf{H}_{k,m}[t]\mathbf{H}_{k,m}^{\text H}[t]{\bf W}_{k,m}[t]\mathbf{N}_{k,m}^{-1}[t], 
\end{equation}
where $\mathbf{N}_{k,m}[t]$ denotes an inter-MU interference plus noise, which is calculated as
\begin{equation}
    \begin{aligned}
        {\mathbf{N}_{k,m}[t]} &{=\!\!\! \sum_{i=1,i\neq k}^{K}\!\sum_{j=1}^{M}\alpha_{i,j}[t]P_{k}[t]{\bf W}_{i,j}^{\text H}[t]\mathbf{H}_{i,j}[t]\mathbf{H}_{i,j}^{\text H}[t]{\bf W}_{i,j}[t]} \\
        &{+\!{\bf W}_{i,j}^{\text H}[t]({\bf \Xi}_{m}\!+\!{\bf \Xi}_{i})\mathbf{W}_{m}[t]\mathbf{W}_{m}^{\text H}[t]({\bf \Xi}_{m}\!+\!{\bf \Xi}_{i})^{\text H}{\bf W}_{i,j}[t]} \\
        & {+\sigma^{2}\mathbf{I}_{W_R}. }
    \end{aligned}
\end{equation}

Then, we denote the total available bandwidth as $B$ and design the achievable communication rate (in bits/s/Hz) from MU $k$ to UAV $m$ as 
\begin{equation}
    R_{k,m}[t]=B\log_{2}\det\left( \mathbf{I}_{W_R}+ \mathbf{\Upsilon}_{k,m}[t]\right).
\end{equation}

\subsection{Radar Sensing Model}
We evaluate the perceptual capabilities of radar using the radar-estimated information rate, which is employed as a performance metric \cite{9840900}. More specifically, the radar signal directed towards a target, in the context of a radar system, can be considered as information about target-related parameters that the target unintentionally transmits, such as the target's reflectivity and its distance from the source. We can regard the radar-estimated information rate as a mutual information amid radar and target, quantifying the amount of valuable target information that can be extracted from the received echoes signal by ISCC devices.

Once the offloading signals from all MUs are decoded, SIC is applied by the receiver to eliminate communication signals from the received waveform, enabling the acquisition of an interference-free radar sensing signal. Hence, the receiver's effective signal for target sensing, denoted as ${\bf c}_{m}[t]\in \mathbb{C}^{W_{R}\times 1}$, is given as
\begin{equation}
    \begin{aligned}
        {\bf c}_{m}^{\text H}[t]{\bf y}_{m,r}[t]&={\bf c}_{m}^{\text H}[t]{\bf \Xi}_{m}\mathbf{W}_{m}[t]{\bf s}_{m}[t]\\ &+{\bf c}_{m}^{\text H}[t]{\bf \Xi}_{i}\mathbf{W}_{m}[t]{\bf s}_{m}[t]+{\bf c}_{m}^{\text H}[t]{\bf z}_{m}[t], 
    \end{aligned}
\end{equation}
which results in the following sensing SINR
\begin{equation}
    \gamma_{m}[t]=\frac{{\left\lVert {\bf c}_{m}^{\text H}[t]{\bf \Xi}_{m}\mathbf{W}_{m}[t]\right\rVert}_{F}^{2}}{ {\bf c}_{m}^{\text H}[t] {\bf R}_{m}[t]{\bf c}_{m}[t]}.
\end{equation}
Here, ${\bf R}_{m}[t]$ denotes the covariance matrix that accounts for both clutter interference and noise, which is given by 
\begin{equation}
    {\bf R}_{m}[t]={\bf \Xi}_{i}\mathbf{W}_{m}[t]\mathbf{W}_{m}^{\text H}[t]{\bf \Xi}_{i}^{\text H}+\sigma^{2}\mathbf{I}_{W_R}.
\end{equation}

We denote $\varrho $ and $\xi $ as the duty factor and pulse duration of the radar, respectively. Consequently, the UAV $m$'s radar estimation information rate can be expressed as 
\begin{equation}
    R_{m}^{\text{rad}}[t]=\frac{\varrho }{2\xi}\log_{2}\left( 1+2B\mu\gamma_{m}[t]\right).
\end{equation}

It is important to mention that the radar-estimated information rate of the radar information determines the range and localization performance of target sensing. To safeguard the performance of radar sensing, we stipulate that the radar-estimated information rate $R_{m}^{\text{rad}}[t]$ must be greater than $R_{\text {rad}}^{\min}$, i.e., $R_{m}^{\text{rad}}[t] \geq R_{\text {rad}}^{\min}$.

\subsection{Computation Model}
We denote the computational task of MU $k$ by a tuple $\Phi _{k}[t]=\left(D_{k}[t],C_{k}[t],J_{k}[t],\beta_{k}[t],T_{k}^{\max}[t]\right)$. Herein, $D_{k}[t]$ refers to the size of data for the computational task, $C_{k}[t]$ and $J_{k}[t]$ denote the mean number of CPU cycles necessary to compute or compress 1 bit of data, respectively. $\beta_{k}[t]$ represents the initial compression ratio at the $k$-th MU, and $T_{k}^{\max}[t]\left( 0\leq T_{k}^{\max}[t]\leq \delta_{t} \right) $ is expressed as the maximum delay. Due to the limited availability of energy and computational resources, accomplishing the task locally within the specified time frame might not be feasible. Therefore, offloading the task to an MEC server for further processing is indispensable. To accomplish this, a partial offloading model that divides the task into two parts is utilized. Defining $\rho_{k}[t]\left( 0\leq \rho_{k}[t]\leq 1 \right)$ as the partition factor, we offload $\rho_{k}[t]$ part of the task data to UAV, while another part, with the data size of $\left(1-\rho_{k}[t]\right)D_{k}[t]$, is local processed. 

In addition, to optimize data storage utilization and enhance data transmission efficiency, data compression integrates into the proposed framework. Similar to the data offloading, the partial data compression model is also considered in our work, where only $\eta_{k}[t]\left( 0\leq \eta_{k}[t]\leq 1 \right)$ part of data is compressed. Furthermore, we express the overall compression ratio of the entire task as $\hat{\beta}_{k}[t]=\eta_{k}[t]\beta_{k}[t]+(1-\eta_{k}[t])$. 

\subsubsection{Local computing and compression} We represent $f_{k}[t]$ as the MU $k$'s computational capacity. In this regard, the delay incurred by MU $k$ in executing its workload is
\begin{equation}
    t_{k}^{\text loc}[t]=\frac{\left(1-\rho_{k}[t]\right)D_{k}[t]C_{k}[t]}{f_{k}[t]}.
\end{equation} 

Furthermore, we express the task data of MU $k$, which needs to be transmitted to UAV $m$ as $\tau_{k}[t]D_{k}[t]$, where $\tau_{k}[t]=\rho_{k}[t]\left( \eta_{k}[t]\beta_{k}[t]+(1-\eta_{k}[t])\right) $. Consequently, the compression latency of MU $k$ can be characterized by
\begin{equation}
    t_{k}^{\text {dc}}[t]=\frac{\rho_{k}[t]\eta_{k}[t]D_{k}[t]J_{k}[t]}{f_{k}[t]}.
\end{equation}

\subsubsection{Computation of floading} After compression, MU $k$ offloads the compressed data to UAV $m$, and the delay associated with the transmission can be expressed as
\begin{equation}
    t_{k}^{\text {off}}[t]=\frac{\tau_{k}[t]D_{k}[t]}{R_{k,m}[t]}.
\end{equation}

When UAV $m$ receives the data offloaded by MU $k$, it first performs the decompression on the compressed data. According to the compression operation described earlier, it can be inferred that the data being decompressed is only $\rho_{k}[t]\eta_{k}[t]D_{k}[t]$. We denote $f_{k,m}^{e}[t]$ as the frequency of MU $k$ allocated by UAV $m$ and the number of CPU cycles required to decompress 1-bit data as $J_{m}[t]$. Thus, the UAV $m$'s decompression latency and computation delay of UAV $m$ for MU $k$ are respectively calculated as
\begin{equation}
    t_{k,m}^{\text {dd}}[t]=\frac{\rho_{k}[t]\eta_{k}[t]D_{k}[t]J_{m}[t]}{f_{k,m}^{e}[t]},
\end{equation}

\begin{equation}
    t_{k,m}^{\text {con}}[t]=\frac{\rho_{k}[t]D_{k}[n]C_{k}[t]}{f_{k,m}^{e}[t]}.
\end{equation}

Compared to the quantity of data transmitted in the uplink, the amount of data, which results from computation, can be omitted, and its impact on latency is neglected in this paper. Hence, the UAV $m$'s total latency for MU $k$ is calculated as 
\begin{equation}
    t_{k,m}^{\text {e}}[t]=t_{k}^{\text {off}}[t]+t_{k}^{\text {dd}}[t]+t_{k,m}^{\text {con}}[t]. 
\end{equation}

\subsection{Energy Consumption Model}
First, we calculate the MUs' energy consumption. Defining $\kappa_{1}$ as the effective capacitance coefficient of MU $k$'s CPU, we can express the MU $k$'s energy consumption during compression, local computing, and transmission are respectively expressed as
\begin{equation}
    e_{k}^{\text{dc}}[t]=\kappa_{1}\rho_{k}[t]\eta_{k}[t]D_{k}[t]J_{k}[t]f_{k}^{2}[t]. 
\end{equation}
\begin{equation}
    e_{k}^{\text{loc}}[t]=\kappa_{1}\left(1-\rho_{k}[t]\right)D_{k}[t]C_{k}[t]f_{k}^{2}[t],
\end{equation}
\begin{equation}
    {e_{k}^{\text{off}}[t]=t_{k}^{\text {off}}[t]P_k[t].}
\end{equation}

Based on the analysis provided above, the MU $k$'s energy consumption yields 
\begin{equation}
    e_{k}[t]=e_{k}^{\text{loc}}[t]+e_{k}^{\text{dc}}[t]+e_{k}^{\text{off}}[t]. 
\end{equation}

In the same way as the effective capacitance coefficient of MU, we define that of UAV $m$'s CPU as $\kappa_{2}$. Thus, when UAV $m$ offers computing services to MUs, the computation energy is written as
\begin{equation}
    e_{k,m}^{\text{con}}[t]=\sum_{k = 1}^{K}\kappa_{2}\alpha_{k,m}[t]f_{k,m}^{e}[t]^{2}\rho_{k}[t]D_{k}[t]C_{k}[t], \label{29}
\end{equation}
and the UAV $m$'s decompression energy is formulated as 
\begin{equation}
    e_{k,m}^{\text{dd}}[t]=\sum_{k = 1}^{K}\kappa_{2}\alpha_{k,m}[t]f_{k,m}^{e}[t]^{2}\rho_{k}[t]\eta_{k}[t]D_{k}[t]J_{m}[t]. \label{30}
\end{equation}

In addition, flying energy is also a significant component of the energy consumption. To prevent collisions among UAVs during task execution, we establish a minimum safety distance $d_{\min }$ between them. We also define the maximum velocity $v_{\max}$ and maximum acceleration $a_{\max}$ of each UAV, which ensure the authenticity of the UAV trajectory. Therefore, UAVs are required to adhere to the following constraints within each time slot:
\begin{equation}
    {\bf q}_m[t+1] = {\bf q}_m[t] + {\bf v}_m[t]\delta_t + \frac{1}{2}{\bf a}_m[t]\delta_t^2, \forall m \in \mathcal{M},t \in \mathcal{T}
\end{equation}
\begin{equation}
    {\left\lVert {\bf q}_i[t]-{\bf q}_j[t]\right\rVert}^2\geq d_{\min }^{2},\forall i,j \in \mathcal{M},i\neq j,t \in \mathcal{T},\label{con:26}
\end{equation}
\begin{equation}
    \left\lVert{\bf a}_m[t]\right\rVert\leq a_{\max}, \left\lVert{\bf v}_m[t]\right\rVert\leq v_{\max},\forall m \in \mathcal{M},t \in \mathcal{T},\label{con:27}
\end{equation}

Then, we denote the flight power of UAV $m$ as
\begin{equation}
    \begin{aligned}
        p_{m}^{\text{fly}}[t]&=P_{1}\left( 1+\frac{3{\left\lVert {\bf v}_m[t]\right\rVert}^{2}}{U_{\text{tip}}^{2}}\right)+\frac{1}{2}d_{0}\rho_{0}sA{\left\lVert {\bf v}_m[t]\right\rVert}^{3} \\ &+P_{2}{\left( \sqrt{1+\frac{{\left\lVert {\bf v}_m[t]\right\rVert}^{4}}{4v_{0}^{2}}}-\frac{{\left\lVert {\bf v}_m[t]\right\rVert}^{2}}{2v_{0}^{2}}  \right)}^{\frac{1}{2}},
    \end{aligned}  
\end{equation}
where $P_{1}$ and $P_{2}$ are the power of UAV's blade and the induced power during hovering, respectively. $U_{\text{tip}}$ represents the blade's tip speed, $d_{0}$ denotes the fuselage drag ratio, and $v_{0}$ is the mean velocity of rotors. Other relevant parameters include the rotor solidity $s$, the air density $\rho_{0}$, and the rotor area $A$ \cite{8663615}. Therefore, we calculate the flight energy of UAV $m$ as 
\begin{equation}
    e_{m}^{\text{fly}}[t]=p_{m}^{\text{fly}}[t]\delta_{t}. \label{35}
\end{equation}

Consequently, the overall energy consumption of UAV $m$ can be computed by combining \eqref{29}, \eqref{30} and \eqref{35}, which is given by 
\begin{equation}
    e_{m}[t]=e_{m}^{\text{fly}}[t]+e_{m}^{\text{dd}}[t]+e_{m}^{\text{con}}[t].
\end{equation}     

\subsection{Problem Formulation}
To tackle the limited energy budget of UAVs and deficient energy resource of MUs, we propose a multi-UAV-assisted ISCC network that leverages MIMO transmission and data compression to jointly optimize the offloading proportion $\boldsymbol{\varrho}\triangleq\{\rho_{k}[t],\forall k\in\mathcal{K},t\in\mathcal{T}\}$, the association factor of MUs $\mathbf{A}\triangleq\left\{ \alpha _{k,m}[t], \forall k\in\mathcal{K},t\in\mathcal{T},m\in\mathcal{M^{*}}\cup\left\{0\right\}\right\} $, the compression ratio of offloaded data $\Pi\triangleq\left\{ \eta_{k}[t], \forall k\in\mathcal{K},t\in\mathcal{T}\right\}$, the CPU frequency of MUs ${\mathbf{F}}_{k}\triangleq\left\{ f_{k}[t], \forall k\in\mathcal{K},t\in\mathcal{T}\right\} $, the computational resource allocation of UAVs ${\mathbf{F}}_{m}\triangleq\{f_{k,m}^{e}[t], \forall k\in\mathcal{K},t\in\mathcal{T},m\in\mathcal{M}\}$, the trajectory planning of UAVs $\mathbf{Q}\triangleq\{ {\bf q}_{m}[t], \forall m\in\!\mathcal{M},t\in\mathcal{T}\}$, the beamforming matrix of radar waveform ${\mathit{W}}_{r}\triangleq\left\{ \mathbf{W}_{m}[t], \forall m\in\mathcal{M},t\in\mathcal{T}\right\}$, and the beamforming matrix of communication symbols ${\mathit{W}}_{c}\triangleq\left\{ \mathbf{W} _{k,m}[t], \forall k \in \mathcal{K},t\in\mathcal{T}, m\in\mathcal{M}\right\}$. Accordingly, the problem of minimizing the weighted energy consumption of both UAVs and MUs, subject to the above-mentioned optimization variables, can be formulated as
\begin{subequations}\label{P:0}
    \begin{align}
        &\mathop {\min }\limits_{ {\varrho},{\mathbf{A}},{\Pi},{{\mathbf{F}}_{k}},{{\mathbf{F}}_{m}},{\mathbf{V}},{{\mathit{W}}_{r}},{{\mathit{W}}_{c}}} \omega \sum_{t = 1}^{T_{\text{ts}}}\sum_{m = 1}^{M}e_{m}[t]+\sum_{t = 1}^{T_{\text{ts}}}\sum_{k = 1}^{K}e_{k}[t]  \label{P:OB}\\
        \text{s.t.}~
        & \eqref{con:1},\eqref{con:2},\eqref{con:26},\eqref{con:27}, \\
        & 0\leq \rho_{k}[t]\leq 1,\forall k \in \mathcal{K},t \in \mathcal{T},\label{con:41c}\\
        & 0\leq \eta_{k}[t]\leq 1,\forall k \in \mathcal{K},t \in \mathcal{T},\label{con:41d}\\
        & 0\leq f_{k}[t]\leq f_{k}^{\max},\forall k \in \mathcal{K},t \in \mathcal{T},\label{con:41e}\\
        & 0\leq f_{k,m}^{e}[t]\leq f_{m}^{\max},\forall k \in \mathcal{K},t \in \mathcal{T},m \in \mathcal{M},\label{con:41f}\\
        &0\leq \sum_{k = 1}^{K}\alpha_{k,m}[t]f_{k,m}^{e}[t]\leq f_{m}^{\max},\forall k \in \mathcal{K},t \in \mathcal{T},m \in \mathcal{M},\label{con:41g} \\
        & 0\leq P_{m}[t]\leq P_{m}^{\max},\forall m \in \mathcal{M},t \in \mathcal{T},\label{con:41h}\\
        & R_{m}^{\text {rad}}[t]\geq R_{\text {rad}}^{\min},\forall m \in \mathcal{M},t \in \mathcal{T},\label{con:41i}\\
        & t_{k}^{\text {dc}}[t]\!+\!\max \{ t_{k}^{\text {loc}}[t],t_{k}^{\text {off}}[t]\!+\!t_{k}^{\text {dd}}[t]\!+\!t_{k,m}^{\text {com}}[t]\}\!\leq\!T_{k}^{\max}[t] , \nonumber \\ 
        &\forall k \in \mathcal{K},t \in \mathcal{T},m \in \mathcal{M},\label{con:41j}  
    \end{align}
\end{subequations}
in which $\varpi $ is the weight factor, $f_{m}^{\max}$ and $f_{k}^{\max}$ are the maximum computational resource of UAV $m$ and  MU $k$, respectively, and $P_{m}^{\max}$ is the maximum transmit power of UAV $m$. Constraints \eqref{con:1} and \eqref{con:2} guarantee the validity of the association status. The minimum safe distance among UAVs is shown in \eqref{con:26}. Constraint \eqref{con:27} ensures the velocity and acceleration of UAVs. Constraint \eqref{con:41c} represents the task-partition factor, while \eqref{con:41d} denotes the task-compression factor. Constraint \eqref{con:41e} denotes the computation resource contraint of MU $k$. Constraint \eqref{con:41f}, concurrently with constraint \eqref{con:41g}, is the computration resource allocation limitation of UAV $m$. The power contraint of UAVs is shown in \eqref{con:41h}. The radar sensing requirement of UAVs is given in \eqref{con:41i}, which denotes that the radar estimation information rate of UAV $m$ cannot be smaller than $R_{\text {rad}}^{\min}$. Constraint \eqref{con:41j} specifies the acceptable computation delay.

\section{PROPOSED DRL APPROACH: ATB-MAPPO}\label{s:proposed}
The optimization problem \eqref{P:0} is a complex mixed integer non-convex problem, characterized by a large number of highly coupled variables. Additionally, the dynamic and uncertain nature of the environment, resulting from time-varying channels and diverse task capabilities, poses challenges for traditional offline optimization methods \cite{9254093}. Aiming at addressing this problem in real-time decision-making, we introduce a DRL approach to configure heterogeneous resources jointly and propose an ATB-MAPPO training framework for the multi-UAV-assisted ISCC network as it enables the involvement of multiple policy types for cooperative and distributed optimization variable decision-making.

\subsection{Multi-agent MDP Model}
First, our problem is expressed as a multi-agent MDP, comprising three essential elements, which includes an action space $\mathcal{A}$, a state space $\mathcal{S}$, and a reward function $\mathcal{R}$. To alleviate the complexity of decision-making for the agents and seek the near-optimal solutions, the agent is decomposed, set as $ u \in \mathcal{U}\triangleq\{1,2,\cdots,U\}$, into two types, which is corresponded to UAVs and MUs. However, in a multi-agent framework, it is difficult for each agent to observe the global state $\mathcal{S}$. Instead, each of them can only obtain a local observation, denoted as $o_{n}^{u}$, in which $n$ denotes the time step. The collective combination of all local observations forms the global state, which can be represented as $\mathcal{S}=\left[\mathcal{O}_{1}, \mathcal{O}_{2},\cdots,\mathcal{O}_{U}\right] $, and the action space can be expressed as $\mathcal{A}=\left[\mathcal{A}_{1},\mathcal{A}_{2},\cdots,\mathcal{A}_{U}\right] $. Hence, two types of agents are described as follows:

\subsubsection{MDP of MU agents} These agents primarily specialize in data compression, task offloading, and local computing configuration for MUs. The set of indices representing these MU agents is denoted as $\mathcal{U}_1\triangleq\{1,2,\cdots,K\} $. Aiming at determining the association, compression proportion, and offloading proportion with UAVs, these agents require observe the tasks, the positions of them, as well as the UAVs' locations.

\textbf{Observation}: The observation for MUs agents is expressed as
\begin{equation}
    o_t^{k}=\{k, \Phi _{k}[t], {\bf q}_m[t], {\bf w}_k[t],  \forall m \in \mathcal{M}\}.
\end{equation}

It is worth noting that each MU can only access its own location, while the coordinates of other MUs remain unknown. In contrast, the positions of all UAVs are known to MUs, as UAVs act as airborne servers. In the case of coordinates with random deviation, we determine the upper and lower bounds on the size of region and scale them into $[0,1]$. The same treatment is similarly applied to address the significant difference in task size. Considering the dynamic voltage frequency scaling technology\cite{9404260},\cite{7542156}, the CPU frequency $\tilde{f}_{k}[t]$ is easily set as $\tilde{f}_{k}[t]=\min\left\{{f}_{k}^{\max},\frac{1}{T_{k}^{\max}[t]}D_k[t]C_k[t]\right\}$ to minimize computational energy consumption.

\textbf{Action}: For the MU agent $k$, its action includes the association factor, the offloading proportion and the compression proportion, and thus can be defined as 
\begin{equation}
    a_t^{k}=\{ \alpha_{k,m}[t], \rho_k[t], \eta_k[t], \forall m \in \mathcal{M}\}.
\end{equation}

\textbf{Reward}: The reward function for MU agents must encompass both the desired objective and penalties associated with failing to  satisfy the latency requirements. Additionally, decomposing the energy consumption between MUs and the associated UAVs is essential. As a result, the reward function for MU agent $k$ is
\begin{equation}
    r_t^{k}=-\left( \omega\sum_{m = 1}^{M}\alpha_{k,m}[t]e_{m}[t] +e_{k}[t]  \right)P_{t,T}^{k}[t]. \label{32}
\end{equation}
We define the function $P(x,\zeta ,\eta )=2-e^{-\left\lceil (x-\zeta )/\eta \right\rceil^{+}} $, the latency penalty thus is computed as 
\begin{equation}
    \begin{aligned}
        P_{t,T}^{k}[t] = &P\left( {\sum_{m = 1}^{M} \alpha_{k,m}[t](t_{k}^{\text {dc}}[t] +\!\max \{ t_{k}^{\text {loc}}[t], } \right. \\
        &\left. { t_{k}^{\text {off}}[t]\!+\!t_{k}^{\text {dd}}[t]\!+\!t_{k,m}^{\text {com}}[t]\}),T_k^{\max}[t],T_k^{\max}[t] } \right). \\
    \end{aligned}
\end{equation}

\subsubsection{MDP of UAV agents} Each UAV is responsible for controlling its speed and allocating the CPU frequency for MUs. Let $\mathcal{U}_2\triangleq\{K+1,K+2,\cdots,K+M\} $ represent the index set of UAV agents. The MDP elements associated with UAV agents are described as follows: 

\textbf{Observation}: The UAVs have access to the positions and task of the paired MUs. Based on this information, we denote $\mathcal{K}_m$ as the set of the MUs matched with UAV $m$ and express the observation as 
\begin{equation}
    \begin{aligned}
        o_t^{K+m}\!\!=\!\!\left\{  {m, {\bf w}_{k}[t],\Phi _{k}[t], {\bf q}_m[t], {\bf q}_{-m}[t], \rho_k[t], \eta_k[t], \forall k \in \mathcal{K}_m} \right\},
    \end{aligned}
\end{equation}
where $-m$ is the indexes in set $\mathcal{M}\backslash\left\{m\right\} $.

\textbf{Action}: In addition to making movement decisions, the UAVs also need to allocate CPU frequencies for  tasks offloaded by MUs. Thus, the actions of the UAVs is defined as
\begin{equation}
    a_t^{K+m}=\{ f_{k,m}^{e}[t], {\bf a}_m[t], \forall k \in \mathcal{K}_{m}\}.
\end{equation}

\textbf{Reward}: For UAV $m$, both the weighted energy consumption and the distance to other UAVs should be heavily considered to optimize the fairness and the channel gain. Additionally, the factors such as collision avoidance, penalties for flying outside designated areas, and insufficient radar perception must be taken into account. Denoting the average energy consumption as $\bar{e}_{m}^{\omega}[t]=\frac{1}{\left\lvert\mathcal{K}_{m}\right\rvert}\sum_{k \in \mathcal{K}_{m}}\alpha_{k,m}[t]e_k[t]+\varpi_{m}e_m[t]$, we can propose the reward as
\begin{equation}
    \begin{aligned}
    r_{t}^{K+m} = &-\left( {k_{1}\bar{e}_{m}^{\omega}[t]+ k_{2}P\left(\lVert{\bf q}_m[t] - \right. } \right. \\
    &\left. { \left. \frac{1}{\left\lvert\mathcal{K}_{m}\right\rvert}\sum_{k = 1}^{K}\alpha _{k,m}[t]{\bf w}_{k}[t]\rVert, d_{th},D \right) } \right)P_{t,T}^{m}P_{t,O}^{m}P_{t,C}^{m}P_{t,R}^{m}, \label{36} \\
    \end{aligned}
\end{equation}
where $k_{1}$ and $k_{2}$ represent the adjusting factors, $d_{th}$ is the threshold which adjusts the distance between UAVs and MUs, and $D$ is the width of square service region. Moreover, the penalties, including latency penalty $P_{t,T}^{m}[t]$, collision penalty $P_{t,C}^{m}[t]$, boundary penalty $P_{t,O}^{m}[t]$, and radar sensing penalty $P_{t,R}^{m}[t]$, are denoted by respectively, where
\begin{equation}
    \begin{aligned}
        P_{t,T}^{m}[t]\!\! = \mathbb{E}_{k \in \mathcal{K}_{m}}  \left[ {P\left( \sum_{m = 1}^{M} \alpha_{k,m}[t] ( t_{k}^{\text {dc}}[t]+\!\max \{ t_{k}^{\text {loc}}[t], \right. } \right. \\
        \left. { \left. t_{k}^{\text {off}}[t]\!+\!t_{k}^{\text {dd}}[t]\!+\!t_{k,m}^{\text {com}}[t]\}),T_k^{\max}[t],T_k^{\max}[t] \right) } \right] \\
    \end{aligned}
\end{equation}
represents the delay penalty of MUs.
\begin{equation}
    P_{t,C}^{m}[t]=\sum_{i = 1,i\neq m}^{M} P\left(\left\lVert {\bf q}_{m}[t]-{\bf q}_{i}[t]\right\rVert,d_{\min},d_{\min}\right)
\end{equation}
indicates the penalty for UAVs that fly within a distance which less than the safe flying distance, and 
\begin{equation}
    P_{t,O}^{m}[t]=1+\frac{1}{\upsilon_{\max}}\left\lVert {\bf q}_{m}[t]-{\text{clip}\left({\bf q}_{m}[t],0,X\right) }\right\rVert
\end{equation}
denotes the penalty incurred when UAV attempts to exceed the boundary.
\begin{equation}
    P_{t,R}^{m}[t]=1+\frac{1}{R_{\text{rad}}^{\min}} \left(\bar{R}_{\text{rad},m}-R_{\text{rad}}^{\min}\right)^{-}
\end{equation}
is the penalty when UAV $m$'s radar estimation information rate is below the minimum threshold $R_{\text{rad}}^{\min}$.

\subsection{MAPPO-based DRL Training Framework}
We propose MAPPO to train the policies in our framework. In this framework, the actor network $\theta_{v}$ outputs the actions, according to the policy $\pi_{v}(a_t|s_t;\theta_v)$ made by $v$-th type of agents, which can be shared among the homogeneous agents, and the critic network $\omega_{v}$ evaluates the quality of actions based on the given reward $\gamma_t$ and state $s_t$ \cite{7542156}.

\begin{figure}[t]
    \centering
    \includegraphics[width=\columnwidth]{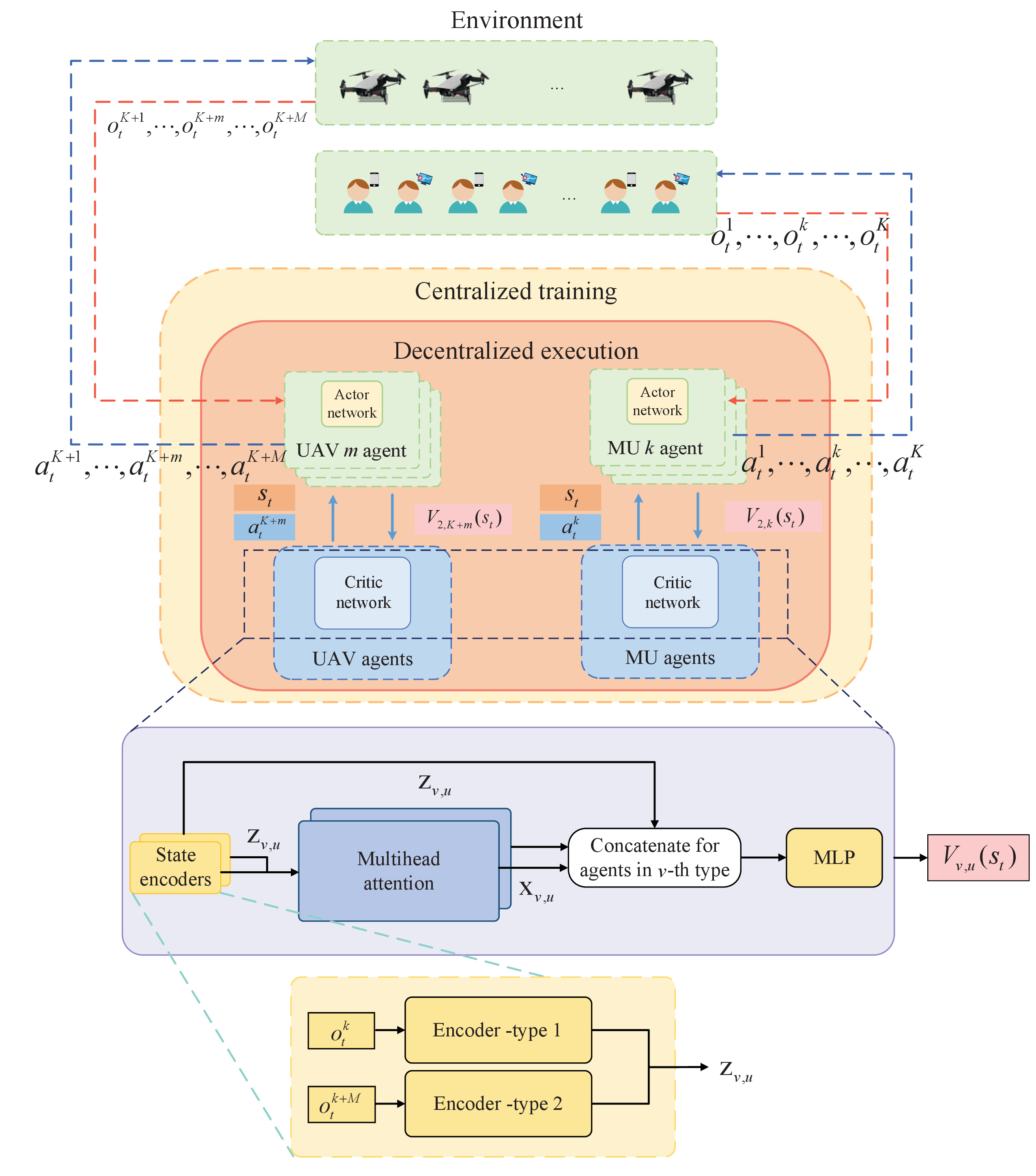}
    \caption{The training framework of MAPPO.}
    \label{fig:MAPPO-framework}
\end{figure}
To facilitate deployment in distributed networks, we employ the centralized training and decentralized executing framework, depicted in Fig. \ref{fig:MAPPO-framework}, which ensures overall performance while giving consideration to the decision-making of each agent. It is noted that the rewards are difficult to be calculated by MUs or UAVs independently during the centralized training process, such that the intervention of a training center is required, where the observations and actions are collected centrally into it to evaluate the rewards. Meanwhile, the training center integrates the observations into the global state and sends them to the critic network of each agent. Then, the critic network generates the state-value network, which is defined as
\begin{equation}
    \begin{split}
        V_{v,t}^{\pi}(s_t,\theta_v)&=\mathbb{E}_{a_t{\sim}\pi_{v}(a_t|s_t;\theta_v)}\left[\mathcal{R}(s_t|a_t),\pi_v\right]\\ &=\mathbb{E}_{a_t{\sim}\pi_{v}(a_t|s_t;\theta_v)}\left[ \sum\limits_{i=0}^\infty\gamma_v^i \mathcal{R}(s_{t+i}|a_{t +i})\mid s_t,\pi_v \right], \label{49}
    \end{split}
\end{equation}
where $\mathbb{E}\left[\cdot\right]$ is the expectation operation, $\mathcal{R}_{u,i}$ represents the reward function for agent $i$ belonging to the $u$-th type of agents, $\gamma_{v}$ denotes reward discount factor and the policy followed by all agents is denoted by $\pi_{v}(a_t|s_t;\theta_v)$. 

From \eqref{49}, it is evident that the state-value function $V_{v,t}^{\pi}(s_t,\theta_v)$ depends on both the current state $s_t$ and the parameters $\theta_v$ of the policy network $\pi_v$. Specifically, as the state $s_t$ improves, the value of $ V_{v,t}^{\pi}$ increases, indicating higher expected rewards. Likewise, when the policy $\pi_v$ with parameters $\theta_v$ performs better, the value of $V_{v,t}^{\pi}$ also becomes higher. When a policy $\pi_v$ consistently achieves outstanding performance for all states $s_t$ within the same time step, the average value of the state-value function $V_{v,t}^{\pi}$ is expected to be significantly higher. Therefore, we can define the objective function for actor network as 
\begin{equation}
    P(\theta_v)=\mathbb{E}_{s_t}\left[V_{v,t}^{\pi}(s_t,\theta_v)\right]. \label{50}
\end{equation}
The objective function is independent of the state $s_t$ and only depends on the parameters $\theta_v$. We adjust the parameters $\theta_v$ of the policy network by using gradient ascent, aiming to maximize the objective function. Let the current parameters of the policy network be denoted as $\theta_v^{\text{now}}$, and after performing the gradient ascent update, we obtain the new parameters $\theta_v^{\text{new}}$ as
\begin{equation}
    \theta_v^{\text{new}}\leftarrow \theta_v^{\text{now}}+\psi \nabla_{\theta_v}P(\theta_v^{\text{now}}),\label{con:j_theta}
\end{equation}
where $\psi$ is the learning rate, and the gradient $\nabla_{\theta_v}P(\theta_v^{\text{now}})$ can be represented as 
\begin{equation}
    \begin{aligned}
        \nabla_{\theta_v}P(\theta_v^{\text{now}})\!&=\!\frac{\partial P(\theta)}{\partial \theta }\bigg|_{\theta =\theta_v^{\text{now}}} \\
    &= \mathbb{E}_{s_t,a_t{\sim}\pi_{v}(a_t|s_t;\theta_v)}\left[ { \nabla_{\theta_v} } \text{ln}\left(\pi_{v}(a_t|s_t;\theta_v)\right) \right. \\
    & \left. { \sum_{l = 1}^{\infty}\left(\gamma_{v}\lambda\right)^{l}\left(Q_{v,u}^\pi (s_t;a_t)-V_{v,u}^{\pi}(s_t,\theta_v)\right) } \right], \label{52} \\
    \end{aligned}
\end{equation}
where $\lambda$ is a balancing parameter, which balances bias and variance of the estimate. We define the action-value function as $Q_{v,u}^\pi (s_t;a_t)$, which is written as 
\begin{equation}
    Q_{v,u}^\pi (s_t;a_t)=\mathbb{E}\left[\sum_{i = 1}^{\infty}\gamma_v^{i}\mathcal{R}_{v,u}(s_{t+i},a_{t+i})\mid s_t=s,a_t=a,\pi_v\right].
\end{equation}

To effectively assess the superiority of an action relative to others and simultaneously reduce variance and bias during the training process, without loss of generality, we utilize generalized advantage estimation in place of the advantage function in \eqref{52}. Similar to the surrogate objective function and the importance sampling in traditional policy optimization. The policy gradient can be redefined as follows:
\begin{equation}
    \begin{aligned}
        \nabla_{\theta_v}P(\theta_v^{\text{now}}) = &\mathbb{E}_{s_t,a_t{\sim}\pi_{v}(a_t|s_t;\theta_v)}\left[ {\frac{\pi_{\theta_v^{\text{new}}}(a_t|s_t)}{\pi_{\theta_v^{\text{now}}}(a_t|s_t)} \cdot} \right. \\
        & \left. { \nabla_{\theta_v} \text{ln}\left(\pi_{v}(a_t|s_t;\theta_v)\right)\hat{A}(s_t) } \right], \label{54}\\
    \end{aligned}
\end{equation}
where $\hat{A}(s_t)=\sum_{l = 1}^{\infty}\left(\gamma_{v}\lambda\right)^{l}\left(r_n+\gamma_vV_v(s_{t+i+1})-V_v(s_t)\right)$.

In addition, to restrict the step size of policy updates and prevent excessive optimization that may lead to instability, we introduce the clip function, where the clip function limits the ratio between the new and old policies, ensuring that the policy updates remain within an appropriate range. The clip function can be expressed as
\begin{equation}
    {\text{clip}\left(x,1-\varsigma ,1+\varsigma \right) =}
    {\left\{
	\begin{aligned}
		& 1+\varsigma,  &\text{if}\ x > 1+\varsigma \\
        & x,  &\text{if}\ 1-\varsigma \leq  x \leq 1+\varsigma \\
		& 1-\varsigma,  &\text{if}\ x < 1-\varsigma \label{55}
	\end{aligned}
	\right.}
\end{equation}
where $\varsigma$ is a regulatory factor. Base on \eqref{50}, \eqref{54} and \eqref{55}, we define $\psi S_{t,u}$ as the policy entropy of the state, thereby, the objective function for actor network, which utilizes the clip function, can be evaluated as
\begin{equation}
    \begin{aligned}
        P(\theta_v) = \mathbb{E}_{s_t}&\left[ { \min\left[ \text{clip}\left(\frac{\pi_{\theta_v^{\text{new}}}(a_t|s_t)}{\pi_{\theta_v^{\text{now}}}(a_t|s_t)},1-\epsilon,1+\epsilon\right)\hat{A}(s_t), \right. } \right. \\
        &\left. { \left. \frac{\pi_{\theta_v^{\text{new}}}(a_t|s_t)}{\pi_{\theta_v^{\text{now}}}(a_t|s_t)} \hat{A}(s_t) \right]+\psi S_{t,v}  } \right]. \\
    \end{aligned}
\end{equation}

Additionally, to update the critic network, which serves as a value network that is utilized to assess the policies of the actor network, we define the loss function of it as
\begin{equation}
    P(\omega _v)= \mathbb{E}_{s_t,a_t}\left[((1-\gamma)\mathcal{R}(s_t|a_t)+\gamma \hat{V}_{\omega _v}(s_{t+1}) - V_{\omega _v}(s_{t}))^2 \right],
\end{equation} 
where $V_{\omega _v}(s_{t})$ is the approximation of the state-value function by the value network with parameter $\omega _v$, $\hat{V}_{\omega _v}(s_{t+1})$ represents the subsequent approximate state when the agent takes action in the present state. Subsequently, we perform gradient descent to update the parameters $\omega _v$ under the learning rate $\varphi$, which is written as
\begin{equation}
    \omega _v^{\text{new}}\leftarrow \omega _v^{\text{now}}-\varphi \nabla_{\omega _v}P(\omega_v^{\text{now}}).\label{con:j_omega}
\end{equation}

\subsection{Beta Distribution and Attention Mechanism}
\subsubsection{Beta policy for actor network} In policy-based DRL algorithms designed for continuous action spaces, it is common to utilize Gaussian distribution, which provides infinite-support probability distribution, for generating output actions. Specifically, in \eqref{49}, the policy of agents adopts a Gaussian policy typically, which can be expressed as
\begin{equation}
    \pi_{v}(a_t|s_t;\theta_v) = \frac{1}{\sqrt{2\pi}\sigma_{\theta_v}(s_t) }\exp \left(-\frac{(a_t - \mu_{\theta_v}(s_t) )^2}{2\sigma_{\theta_v}^2(s_t)} \right) 
\end{equation}
where the mean $\mu_{\theta_v}(s_t)$ and the standard deviation $\sigma_{\theta_v}(s_t)$ are given by a function approximator parameterized by $\theta_v$.
However, by contrast to the Gaussian distribution, the actual action space is bounded due to physical constraints in the real world. This inevitably leads to estimation bias, thus slowing down the training progress and increasing the difficulty of training.

To tackle this issue, the Beta distribution, a finite-support distribution, is adopted as an approach to the issue, which caused by the Gaussian distribution\cite{Chou2017ImprovingSP}. Specifically, we adopt the Beta distribution with the shape parameters $\zeta $ and $\eta $ to define the upper and lower bounds of the action outputs. This guarantees that it does not suffer from boundary effects and has a faster convergence rate owing to unbiasedness. The Beta distribution can be defined as
\begin{equation}
    f(x,\zeta  ,\eta  )=\frac{\Gamma(\zeta +\eta )}{\Gamma(\zeta )\Gamma(\eta )}(1-x)^{\eta -1}x^{\zeta -1},
\end{equation}
where $\Gamma\left(x \right)$ denotes the factorial function of $x-1$. $\zeta -1$ and $\eta -1$ indicate the quantity of successes and failures in the prior knowledge, respectively.

We use $\pi_{v}(a_t|s_t;\theta_v) = f(\frac{a_t + h}{2h} ,\zeta,\eta)$ to represent the stochastic policy and call it the Beta Policy. Since the beta distribution has finite support and no probability density falls outside the boundary, the Beta policy is bias-free. The shape parameters $\zeta = \zeta_{\theta_v}(s_t)$, $\eta = \eta_{\theta_v}(s_t)$ are also modeled by neural networks with parameter $\theta_v$. In this paper, we only consider the case where $\zeta, \eta > 1$, in which the Beta distribution is concave and unimodal.

\subsubsection{Attention mechanism for critic network} As the critic network, its role is to integrate the observations from the agents, calculate the temporal-difference error, and drive the agents to make optimal policies. Nonetheless, as the number of agents increases, employing typical fully connected networks in the critic network becomes challenging to cope with the rapidly escalating complexity, which causes its convergence to be slow or even difficult, thereby affecting the training speed of the agents.

To address this issue, the attention mechanism is incorporated into the critic network \cite{9583941}. The attention mechanism employs a query-key-value model, where each agent queries the observation-action information of other agents and uses it as input to its own critic network. For the $v$-th type of agent $u$, it takes all observation-action information from other agents and feeds it into the Multi-Layer Perceptron (MLP) encoders to extract feature vectors, represented as $\left\{{\bf z}_{v,u},\forall u\in \mathcal{U}\right\} $, which are utilized by attention units to calculate attention values.We define the feature vector of agent $u$ as $\emph{query}$ and that of others are $\emph{value}$ and $\emph{key}$, and the scaling matrices for $\emph{query}$, $\emph{key}$, and $\emph{value}$ are treated as ${\bf W}_{\text {que}}$, ${\bf W}_{\text {key}}$ and ${\bf W}_{\text {val}}$, respectively. The softmax function is utilized for normalization, which generates the attention weights $\iota _{v,w}$ as 
\begin{equation}
    \iota _{v,w}=\!\!\text{Softmax}\!\!\left(\!\frac{{\bf z}_{v,w}^{\text{T}}{\bf W}^{\text{T}}_{\text {key}}{\bf W}_{\text {que}}{\bf z}_{v,u}}{\sum_{i \neq  w} \left({\bf z}_{v,i}^{\text{T}}{\bf W}^{\text{T}}_{\text {key}}{\bf W}_{\text {que}}{\bf z}_{v,u}\right) } \!\right), \forall w \in \mathcal{U}\backslash\!\!\left\{u\right\},
\end{equation}
and the attention value for agent $u$ is obtained by taking the sum of $\emph{value}$, which is same as $\emph{key}$, using the attention weights $\iota _{v,w}$, expressed as ${\bf E}_{v,u}=\sum_{i\neq u}\iota _{v,w}{\bf W}_{\text {val}}{\bf z}_{v,i}$.

According to the attention mechanism, the state-value function $V_{\omega _v}(s_{t})$ output by the critic network can be expressed as
\begin{equation}
    V_{\omega _v}(s_{t})=f_{u}(\omega _v, \text{concat}({\bf E}_{u})),
\end{equation}
where the concat function concatenates all the vector of attention values obtained from multiple attention units into a single vector and $f_{u}$ is the MLP that takes the concatenated attention value and its feature as input, and outputs the final state-value.

In accordance with the aforementioned discussions, the ATB-MAPPO training framework is summarized in Algorithm 1.
\begin{algorithm}[h]
    \caption{Proposed ATB-MAPPO training framework}
    \label{alg:ATB-MAPPO}
    \begin{algorithmic}[1]
        \STATE{Initialize $t=1$ maximum training episodes $\text{Mte}$, PPO epochs $\text{Pec}$, and episode length $\text{Epl}$.}
        \STATE{Initialize the parameters of actor network $\theta_u$ and the parameters of critic network $\omega_u$ of UAVs and MUs, $\forall e \in \left\{1,2\right\} $.}
        \FOR{Episode=1 to Mte}
            \FOR{t=1 to Epl}
                \STATE{The agents of MUs acquire observations $o_t^u$ from the environment, $\forall u \in \mathcal{U}_1$;}
                \STATE{The agents of MUs execute actions $a_t^u$, $\forall u \in \mathcal{U}_1$;}
                \STATE{The agents of UAVs acquire observations $o_t^u$ from the environment, $\forall u \in \mathcal{U}_2$;}
                \STATE{The agents of MU execute actions $a_t^u$, $\forall u \in \mathcal{U}_2$;}
                \STATE{The UAVs and MUs transmit their actions and observations to the control center, where the rewards $r_t^u$ are computed;}
            \ENDFOR
                \STATE{Calculate log-probability $pr_t^u, \forall u \in \mathcal{U}, t \in \{ 1,\cdots, {\rm{Epl}} \}$;}
                \STATE{Summarize the transitions ${\rm{Te}}_t^u = \{ o_t^u,a_t^u,r_t^u,s\left( t \right),pr_t^u,\forall u \in \mathcal{U},t \in \{ 1,\cdots,{\rm{Epl}} \} \}$;}
                \FOR{epoch = 1 to Pec}
                    \FOR{ agents $u \in \mathcal{U}$}
                        \STATE{Adjust $\theta_u$ and $\omega_u$ according to \eqref{con:j_theta} and \eqref{con:j_omega};}
                    \ENDFOR
                \ENDFOR
        \ENDFOR
    \end{algorithmic}
\end{algorithm}

\subsection{Complexity Analysis}
The computational complexity, associated with the ATB-MAPPO algorithm, primarily consists of two parts. On the one hand, the computational complexity generated by the $i$-th layer of the MLP can be expressed as $\mathcal{O} (C_{i-1}C_{i}+C_{i}C_{i+1})$, where $C_{i}$ represents the number of neurons in the $i$-th layer. Denoting the number of the layer of one MLP as $I$, we can calculate the computational complexity of an MLP as $\mathcal{O} (\sum_{i = 2}^{I-1} (C_{i-1}C_{i}+C_{i}C_{i+1}))$. On the other hand, the complexity generated by the attention module can be expressed as $\mathcal{O}(I^2 V)$, in which $V$ denotes the length of feature values output from the state encoders. In our framework, the actor networks are composed of a single MLP, while the critic networks consist of one MLP for value output and two encoders catering to the distinct agent types. Therefore, the computational complexity of the training algorithm for all $\text{Mte}$ episodes is calculated as $\mathcal{O}\left({\text{Mte}}\left({\text{Pec}}\cdot I^2 V+{\text{Epl}}\sum_{i = 2}^{I-1} (C_{i-1}C_{i}+C_{i}C_{i+1})\right) \right) $.

As a comparison, a complexity analysis is performed for the scheme of \cite{9996408}, that considers a similar scenario to this paper, ang uses a WMMSE-based algorithm to optimize resource allocation as well as sensing and communication beamforming. \cite{9791349} considers the design of a beamforming framework in which a convolutional neural network (CNN) module based on learning algorithms is employed, and a convolutional long short-term memory (LSTM) network is used to improve the communication efficiency, which is also comparatively applied to the scenario of this paper, referred to as CNN-LSTM-Net. In addition, an SCA algorithm based beamforming design and resource allocation method in \cite{10158711} is also comparatively applied. Comparison of the complexity of the four algorithms is presented in Table \ref{tab:computational-complexity}. We consider the number of users $K=10$ and the number of antennas $W=8$. Based on the fact that the learning algorithm involves convolutional layers so we set the maximum number of iterations $\text{Mte}=200$ and the intermediate convolutional layers as 2. Although each agent complexity of CNN-LSTM-Net is comparable to the algorithm proposed in this paper, its complexity within the whole iteration is still about 20 times of the proposed algorithm due to the centralised training and centralised execution of the algorithm. The complexity of the WMMSE-based algorithm and the complexity of the SCA-based algorithm are 500 and $\num{6.4e4}$ times higher, respectively, compared to the proposed algorithm. It is obvious that the proposed algorithm exhibits superior complexity advantage.
\begin{table}[t]
    \newcommand{\tabincell}[2]{\begin{tabular}{@{}#1@{}}#2\end{tabular}}  
    \caption{comparison of computational complexity of different algorithms}
    \centering
    \renewcommand{\arraystretch}{1.5}
    \begin{tabular}{p{1.0in}<{\centering} p{2.2in}<{\centering}}
    \toprule[1.5pt]
    Algorithm & Computational complexity \\
    \midrule
    the WMMMSE-based algorithm \cite{9996408} & \tabincell{c}{$\mathcal{O}({\text{Mte}}(2K^2W^3+2K^3$ \\ $+K^{1/2}(4K+W)(3K+W)^2+6K^2))$} \\
    \cline{2-2}
    the CNN-LSTM-Net algorithm \cite{9791349} & \tabincell{c}{$\mathcal{O}({\text{Mte}}\cdot N_e({\text{Epl}}\cdot K\sum_{i = 1}^{I} (C_{i-1}C_{i}a_ib_is_i^2) $\\$ +4{\text{Epl}}(\mathcal{G}_1\mathcal{G}_2+\mathcal{G}_2^2+\mathcal{G}_2)) )$} \\
    \cline{2-2}
    the SCA-based algorithm \cite{10158711} & \tabincell{c}{$\mathcal{O}\left({\text{Mte}}\sqrt{WL+K}(W^6K^3+W^4L^2K+K^3) \right)$}  \\
    \cline{2-2}
    the proposed algorithm & \tabincell{c}{$\mathcal{O}({\text{Mte}}({\text{Pec}}\cdot I^2 V+{\text{Epl}}\sum_{i = 2}^{I-1} (C_{i-1}C_{i} $ \\ $ +C_{i}C_{i+1})) )$}  \\
    \bottomrule[1.5pt]
    \end{tabular}
    \label{tab:computational-complexity}
\end{table}

\section{Simulation Results}\label{s:simulation}
Extensive simulations are conducted in this section to assess the effectiveness of the proposed ATB-MAPPO training framework in the UAV-assisted ISCC network. Initially, we demonstrate the training convergence of the ATB-MAPPO algorithm. Subsequently, we compare the performance of it against following benchmark algorithms utilized in existing research:
\begin{enumerate}
    \item \textbf{Beta-MAPPO}: This benchmark employs the proposed MAPPO-based training algorithm with a Beta distribution on the actor network, omitting the utilization of an attention mechanism.
    \item \textbf{Pure-MAPPO}: This benchmark adopts the proposed MAPPO-based training algorithm, utilizing the widely adopted Gaussian distribution, while excluding the incorporation of an attention mechanism.
    \item \textbf{MADDPG}: This benchmark employs the multi-agent deep deterministic policy gradient (MADDPG) algorithm, which incorporates deterministic action outputs and introduces noise for exploration purposes \cite{9209079}. Each agent in MADDPG corresponds to two shared actor and two critic networks.
    \item \textbf{Energy Minimization}: Each UAV aims to minimize the total energy consumption of all sensing tasks while ensuring the accuracy requirements of the sensing tasks and the minimum energy requirements of the MUs.
    \item \textbf{Without Computation}: Each UAV serves communication MUs and senses targets but does not perform the computation process on the sensing data.
    \item \textbf{Accuracy Maximization}: The goal for each UAV is to maximize the sum of $R_{m}^{\text{rad}}$ to enhance the accuracy of all sensing tasks while ensuring that the energy and delay requirements for the sensing tasks.
\end{enumerate}

\subsection{Simulation Settings}
In our simulated scenarios, we consider a square area spanning 1000 m by 1000 m, where UAVs are uniformly situated at an altitude of 200 m. The MUs are distributed randomly throughout this region, with their coordinates denoted as $x$ and $y$, ranging from 0 to 1000 m. The latency requirements for each task fall within the range of $\left[0.7\text{s},1.0\text{s}\right]$. The size of the task data follows a uniform distribution in the range of $\left[0.5 \text{{Mb}}, D_{\max}\right]$, where $D_{\max}$ is set as the default value of 1.5 Mb \cite{8807184}. Additionally, the average number of cycles needed for 1-bit of compression and computation is specified as $J_k[t]\in \left[100,300\right]$ cycles and $C_k[t]\in \left[500,1500\right] $ cycles, respectively. Other simulation parameters and hyperparameters of the MAPPO algorithm are presented in Table \ref{tab:environment-settings} and Table \ref{tab:algorithm-parameters} \cite{10021296, 8956055}, respectively.
\begin{table}[t]
    \caption{Environment Settings}\centering
    \renewcommand{\arraystretch}{1.2}
    \begin{tabular}{|p{2.3in}<{\centering}|p{0.8in}<{\centering}|}
    \hline
    Parameter & Value \\
    \hline 
    The time period $T$ & 200 s \\
    \hline
    The channel power gain $\rho $& -30 dB \\
    \hline
    The channel bandwidth $B$ & 10 MHz \\
    \hline
    The Rician factor & 10 \\
    \hline
    The time slot $\delta_t$ & 1 s  \\
    \hline
    The noise power & -65 dBm \\
    \hline
    The radar duty factor $\delta $ & 0.01  \\
    \hline
    The radar pulse duration $\mu $ & 2 $\times 10^{-5}$ s \\
    \hline
    The maximum power of MUs $P_{\max}$ & 0.5 W  \\
    \hline
    The minimum radar estimation information rate $R_{\min}$  & 2.2 $\times 10^{4}$ bps\\
    \hline
    The number of antennas for MUs and UAVs $W_r$, $W_t$ & 4  \\
    \hline
    The capacitance coefficient $\kappa_1$ and $\kappa_2$ & $10^{-27}$\\
    \hline
    The maximum CPU frequency of MUs $f_k^{\max}$ & 1 GHz  \\
    \hline
    The maximum acceleration of UAVs $a_{\max}$ & 5 $m/s^2$\\
    \hline
    The maximum CPU frequency of UAVs $f_e^{\max}$ & 10 GHz  \\
    \hline
    The maximum velocity of UAVs $v_{\max}$ & 20 m/s \\
    \hline
    The UAV settings A, $v_{0}$, $U_{\text{{tip}}}$& 0.5030 $m^2$ 3.6 m/s 120 m/s \\
    \hline
    The UAV settings $P_{1}$, $P_{2}$& 59.03\! W 79.07\! W \\
    \hline
    The weight factor $\varpi$ & 0.001  \\
    \hline
    The safe distance between UAVs $d_{\min}$ & 3 m \\
    \hline
    \end{tabular}
    \label{tab:environment-settings}
\end{table}
\begin{table}[h]
    \caption{hyperparameters parameters of algorithms}\centering
    \renewcommand{\arraystretch}{1.2}
    \begin{tabular}{|p{2.3in}<{\centering}|p{0.8in}<{\centering}|}
    \hline
    Parameter & Value \\
    \hline
    The episode length $\text{{Epl}}$ & 200 \\
    \hline
    The maximum training episodes $\text{{Mte}}$ & 300 \\
    \hline
    The discount factor  & 0.98  \\
    \hline
    The penalty factors $\mu_o$,$\mu_t$ and $\mu_c$ & 0.1 \\
    \hline
    The learning rate of actor & 0.0005 \\
    \hline
    The distance threshold & 350 m \\
    \hline
    The adjusting factor $k_1$ and $k_2$ & 0.3 0.7  \\
    \hline
    The number of attention heads $\psi$ & 4 \\
    \hline
    The optimizer & Adam  \\
    \hline
    The length of feature values & 64 \\
    \hline
    The sizes of hidden layers  & 64 and 128 \\
    \hline
    The number of hidden layers for MLP & 2 \\
    \hline
    \end{tabular}
    \label{tab:algorithm-parameters}
\end{table}

\subsection{Performance Evaluation}
\begin{figure}[t]
    \centering 
    \includegraphics[width=\linewidth,scale=1.00]{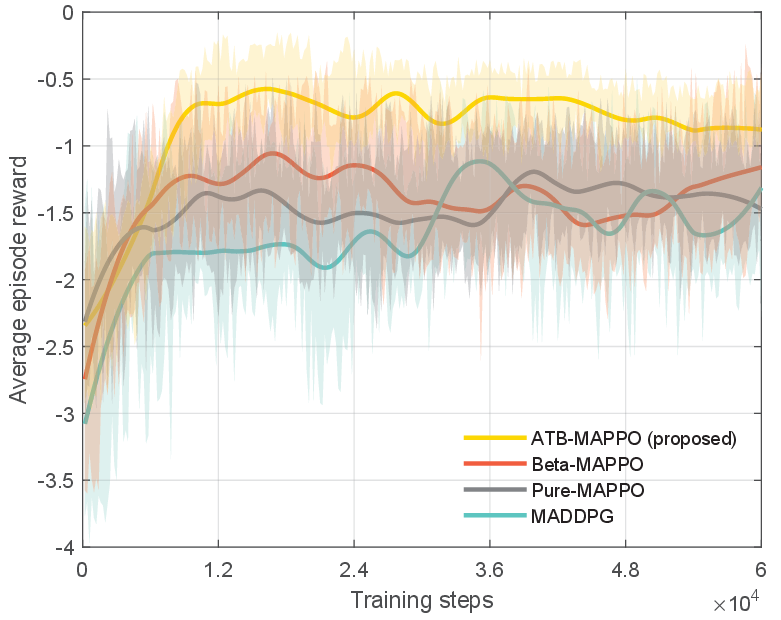}
    \caption{The convergence versus MU agents.}
    \label{fig:MU-reward}
\end{figure}

\begin{figure}[t]
    \centering
    \includegraphics[width=\linewidth,scale=1.00]{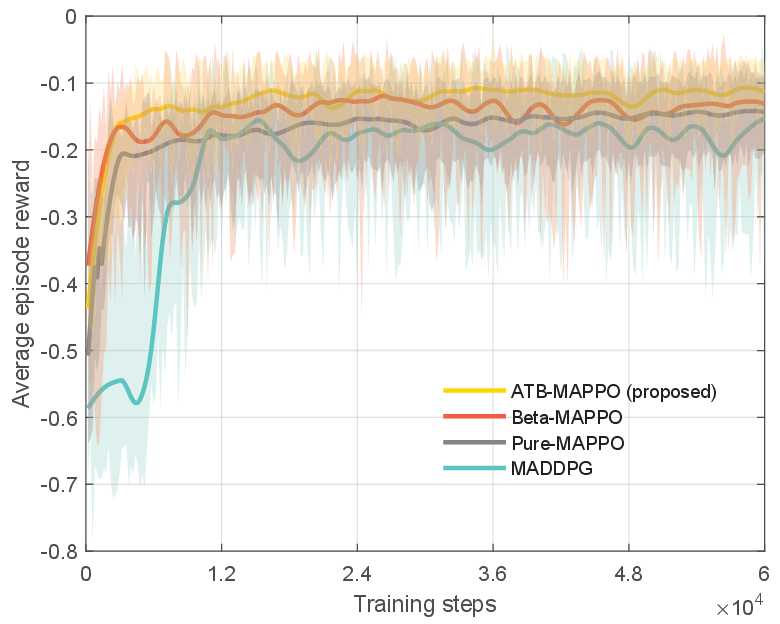}
    \caption{The convergence versus UAV agents.}
    \label{fig:UAV-reward}
\end{figure}

We conduct a comparison between the proposed ATB-MAPPO scheme and other MADRL benchmarks in Fig. \ref{fig:MU-reward} and Fig. \ref{fig:UAV-reward}, aiming at evaluating the convergence behavior, which consider a scenario with $M = 5$ UAVs and $K = 25$ MUs. As depicted in the figures, it is evident that with the increase of training steps, the rewards of all schemes gradually ascend, validating the effectiveness of MADRL algorithms in the context of computation offloading. Notably, the proposed ATB-MAPPO scheme exhibits a faster convergence rate than Beta-MAPPO, and achieves higher rewards than Pure-MAPPO, which signifies the superiority of utilizing the Beta distribution for actions within the proposed ISCC network. Furthermore, compared to the MADDPG algorithm with the off-policy approach, the MAPPO algorithm with the on-policy approach, where the behavior policy aligns with the target policy, can exhibit significantly improved convergence speed and rewards. Based on the on-policy approach, the agents trained by the MAPPO algorithm can more rapidly improve their policies while interacting with the environment, making it easier to adapt to environmental changes while maintaining exploratory behavior. Coupled with the fact that MAPPO has a central controller that provides the intelligences with a global information about the environment, the complex tasks with highly correlated agents can be handled more stably using the MAPPO algorithm, which in turn achieves better convergence. 

\begin{figure}[t]
    \centering 
    \includegraphics[width=\linewidth,scale=1.00]{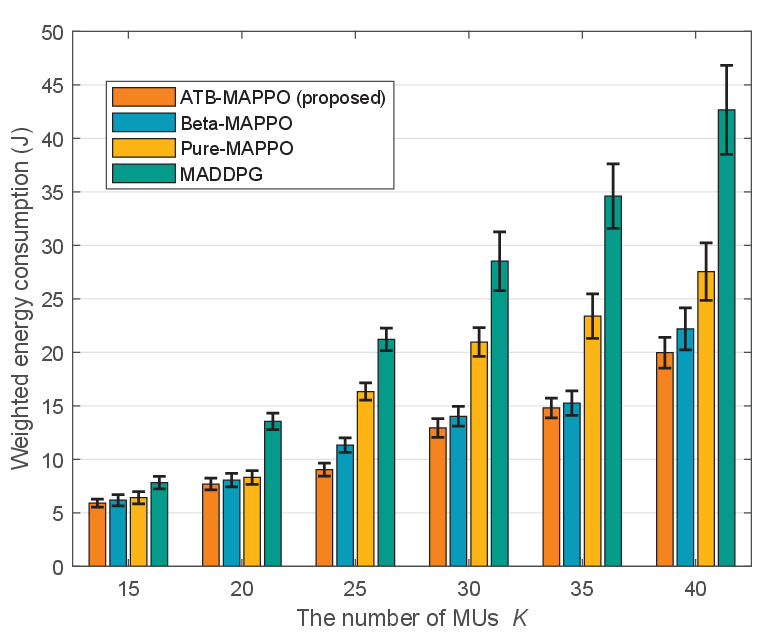} 
    \caption{The performance comparison versus different numbers of MUs.}
    \label{fig:MUs_energy_consumption}
\end{figure}

Fig. \ref{fig:MUs_energy_consumption} illustrates a comprehensive comparison of the weighted energy consumption across different scenarios involving varying numbers of MUs, while considering a fixed number of 5 UAVs. As observed from the results in Fig. \ref{fig:MUs_energy_consumption}, an increase in the number of MUs brings about a corresponding augmentation in the weighted energy consumption of both UAVs and MUs. Notably, the proposed ATB-MAPPO scheme consistently demonstrates superior performance, outperforming both the MADDPG and MAPPO-based schemes by a significant margin. In addition, as more MUs are added to the network, the performance gap  between adjacent settings tends to widen, which is attributed to the elevated signal interference between devices caused by the escalating quantity of MUs, leading to reducing the MU-UAV transmission rate. This inevitably results in higher costs in transmission. To meet the latency requirements, more resources are allocated to computation under the constraint of squeezed computation time.

\begin{figure}[t]
    \centering 
    \includegraphics[width=\linewidth,scale=1.00]{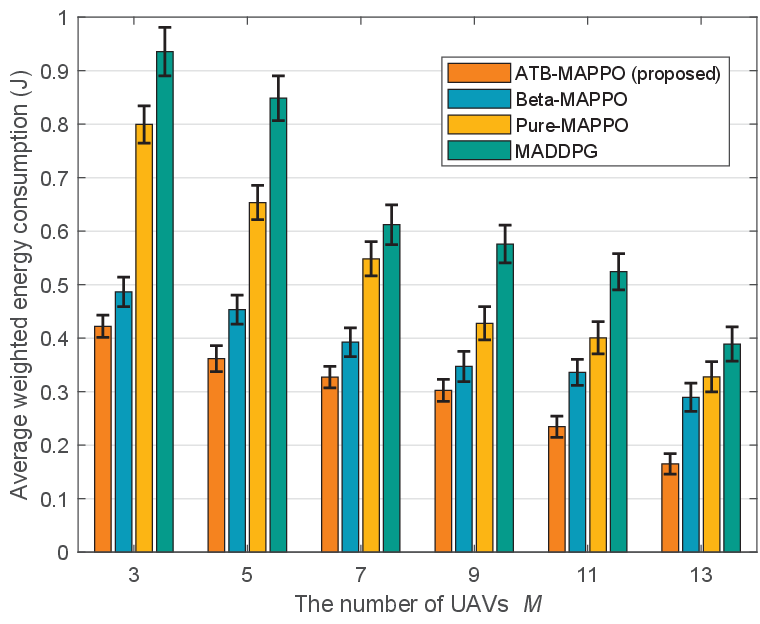}
    \caption{The performance comparison versus different numbers of UAVs.}
    \label{fig:UAVs_energy_consumption}
\end{figure}

Fig. \ref{fig:UAVs_energy_consumption} presents a comparative analysis of the four schemes for varying numbers of UAVs with $K = 25$ MUs. With the growing number of UAVs, a noticeable decrease in the average weighted energy consumption of MUs becomes evident, which indicates that more UAVs contributes to greater computational resource. As training progresses, the computational load between UAVs and MUs is gradually balanced, resulting in decreased energy consumption. Additionally, the performance gap between the MADDPG and MAPPO-based schemes gradually narrows with an increasing number of UAVs, where the proposed scheme consistently outperforms the benchmark schemes. These findings underscore the effectiveness of the proposed algorithm in optimizing policies.

\begin{figure}[t]
    \centering 
    \includegraphics[width=\linewidth,scale=1.00]{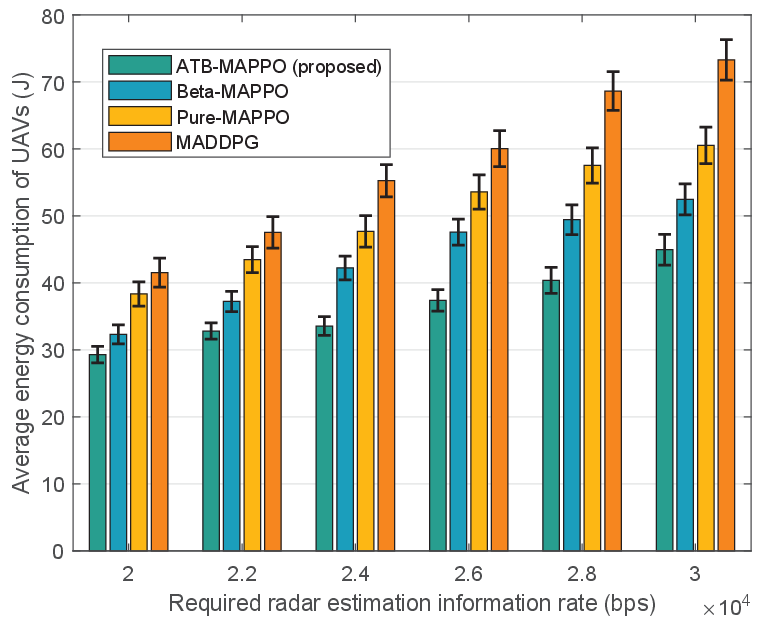} 
    \caption{The performance versus different required estimation rate.}
    \label{fig:R_rad_UAV_energy_consumption}
\end{figure}

\begin{figure}[t]
    \centering 
    \includegraphics[width=\linewidth,scale=1.00]{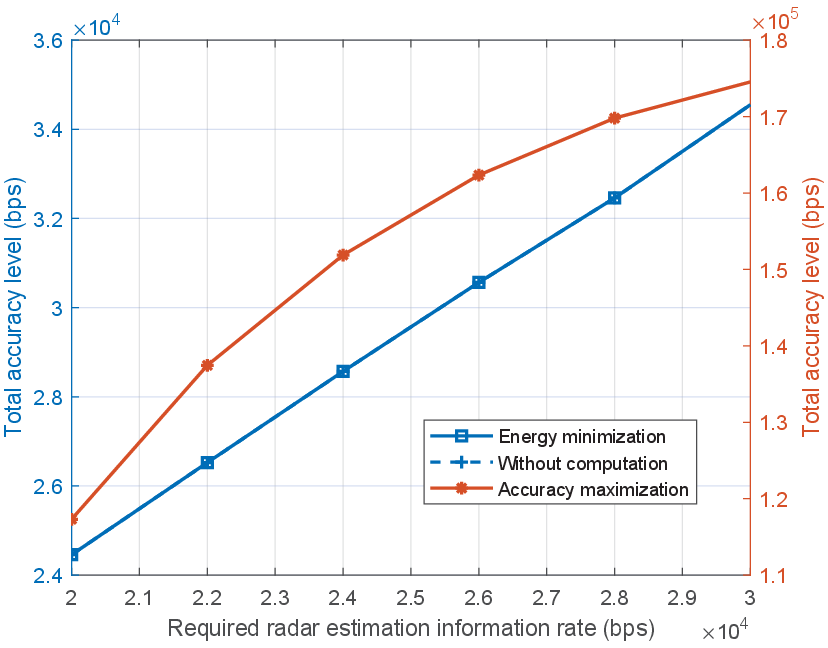} 
    \caption{Total accuracy level of different schemes versus different minimum accuracy thresholds.}
    \label{fig:Total_accuracy_level}
\end{figure}

\begin{figure}[t]
    \centering 
    \includegraphics[width=\linewidth,scale=1.00]{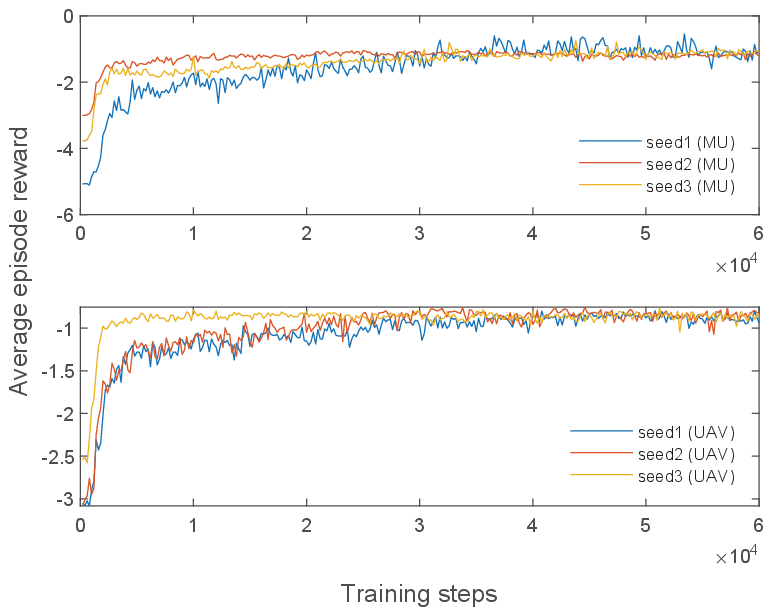}
    \caption{The convergence versus different seeds.}
    \label{fig:seeds}
\end{figure}

The average weighted energy consumption of UAVs, as a function of the required radar estimation information rate, is illustrated in Fig. \ref{fig:R_rad_UAV_energy_consumption}. Overall, the results demonstrate that the average weighted energy consumption increases with an increase in the required estimation information rate, which is in line with our intuitive expectations. Besides, the proposed ATB-MAPPO scheme exhibits superior performance, maintaining a significant performance gap compared to MAPPO-based and MADDPG schemes. This phenomenon can be explained by the fact that an growth in the minimum required estimation information rate results in UAVs spending more time on sensing tasks during interactions with MUs, thereby significantly exacerbating the energy consumption of UAV flight.

Fig. \ref{fig:Total_accuracy_level} illustrates the relationship between the total accuracy level of various schemes and different minimum accuracy thresholds. For all schemes, the total accuracy level increases as the minimum accuracy threshold rises. Across all minimum accuracy threshold settings, the accuracy maximization scheme consistently achieves a higher total accuracy level compared to the other two schemes. This advantage arises because, in the accuracy maximization scheme, UAVs allocate the highest sensing power and position themselves closest to the target, thereby maximizing the total accuracy. In the other two schemes, the total accuracy level is equivalent and equal to the sum of the minimum accuracy thresholds for all sensing tasks, which is due to the fact that the total delay is minimized only when the UAV positioning and power allocation precisely meet the minimum accuracy thresholds.

\begin{figure}[t]
    \centering 
    \includegraphics[width=\linewidth,scale=1.00]{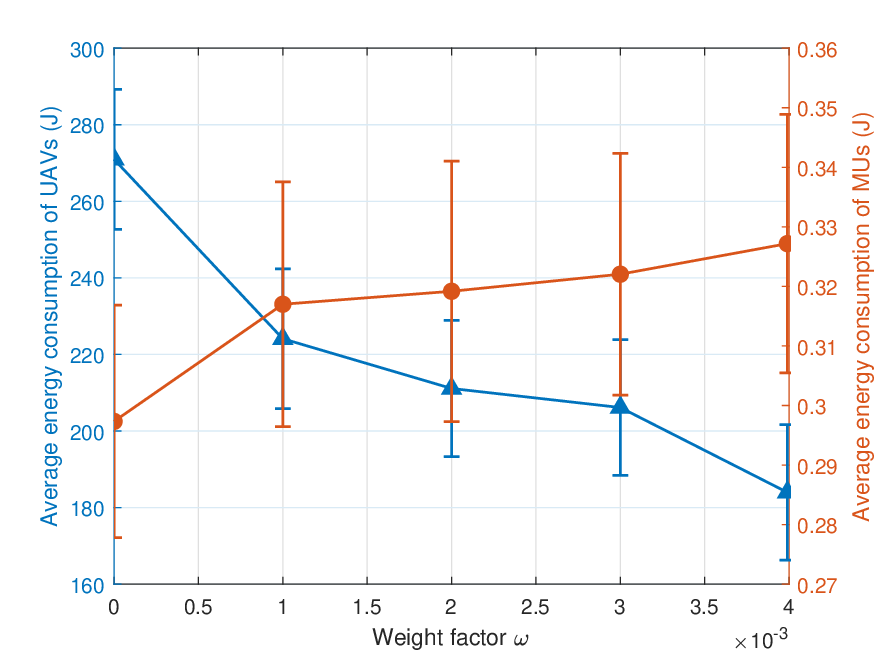}
    \caption{The performance versus different required weight factor.}
    \label{fig:Weight_factor}
\end{figure}

Fig. \ref{fig:Weight_factor} illustrates the relationship between the energy consumption of MUs and UAVs for different weight factors $\omega$. It is evident that an increase in $\omega$ leads to a gradual rise in the energy consumption of MUs, while the energy consumption of UAVs decreases. The variation of $\omega $ on objective function directly impacts the relative significance of MUs and UAVs energy consumption. As evident from \eqref{32} and \eqref{36}, the changes in relative significance has a significant influence on the reward function, thereby altering the polices of agents.

In general, generalization capability is a crucial metric for DRL. In Fig. \ref{fig:seeds}, we compare the convergence of MU and UAV under different training seeds. It is evident that the proposed algorithm achieves rapid convergence for both UAV and MU when faced with entirely new training data. Additionally, we observe that an initial discrepancy in reward values for both MU and UAV is observed when opting for various training seeds, while once convergence is achieved, the reward values stabilize within similar ranges. This initial disparity is attributed to the influence of different training data on the critic network's evaluation of the agent's actions, thus affecting the reward values. However, as the agent continues training, the critic network gradually adjusts its evaluation criteria, leading the agent to choose actions that result in higher reward values. Ultimately, the agents with different seeds converge within the same range.

\begin{figure}[t]
	\centering  
	\subfigure[Scenarios 1: The starting positions of UAVs are random with $M$ = 3 UAVs and $K$ = 15 MUs.]{
		\includegraphics[width=0.48\linewidth]{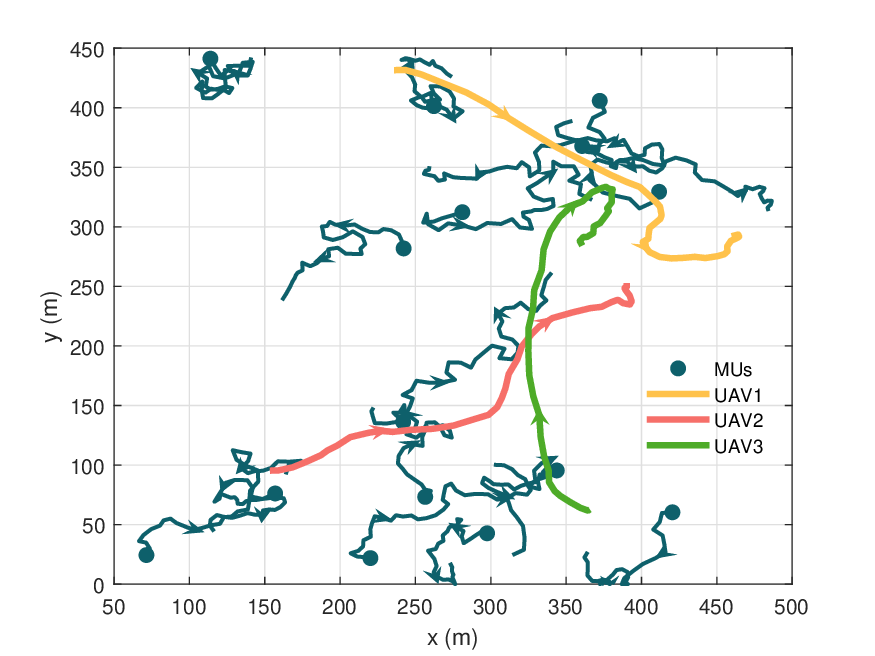}}
	\subfigure[Scenarios 2: The starting positions of UAVs are fixed with $M$ = 3 UAVs and $K$ = 15 MUs. ]{
		\includegraphics[width=0.48\linewidth]{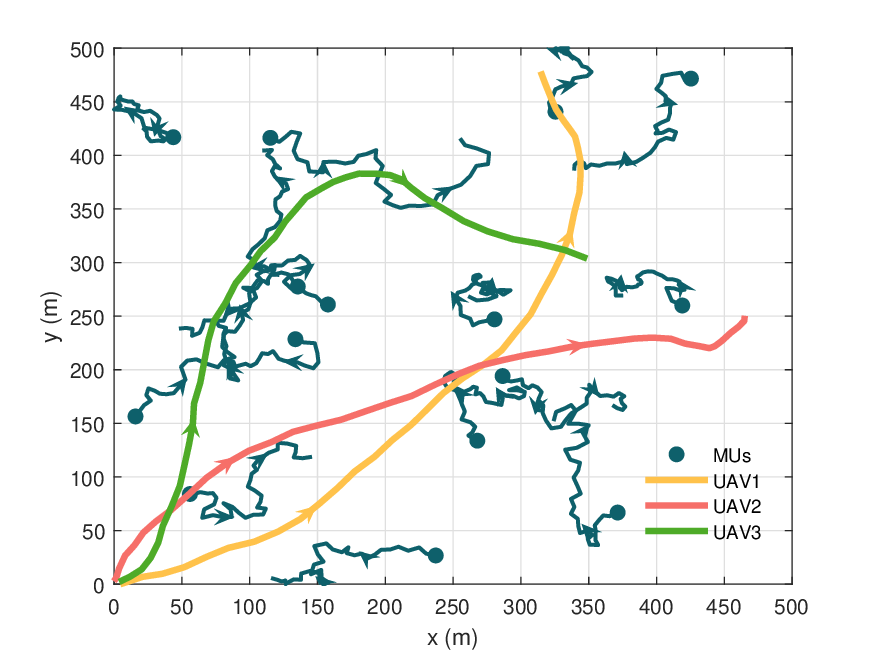}}
    \subfigure[Scenarios 3: The starting positions of UAVs are random with $M$ = 10 UAVs and $K$ = 60 MUs.]{
		\includegraphics[width=0.48\linewidth]{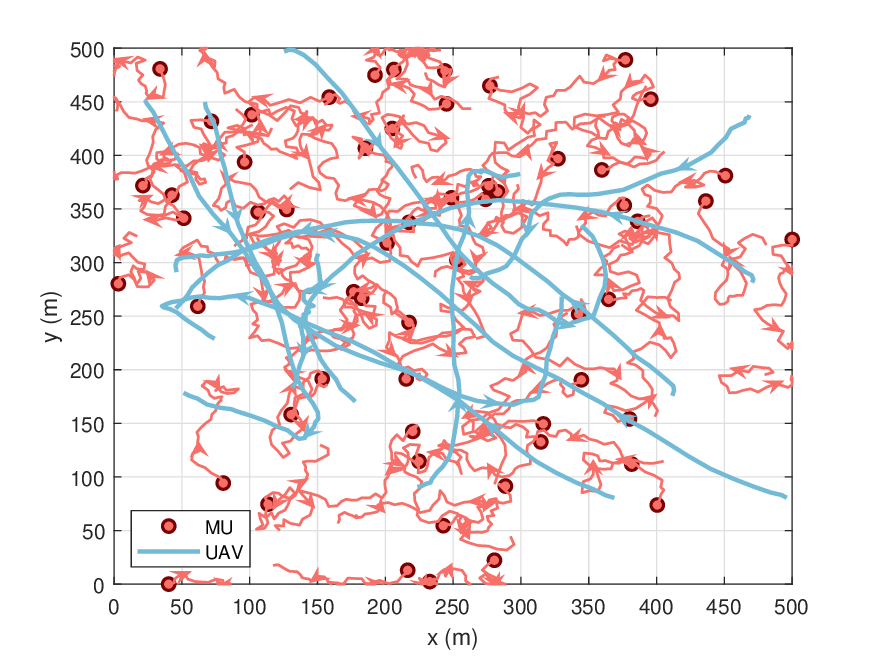}}
	\subfigure[Scenarios 4: The starting positions of UAVs are fixed with $M$ = 10 UAVs and $K$ = 60 MUs.]{
		\includegraphics[width=0.48\linewidth]{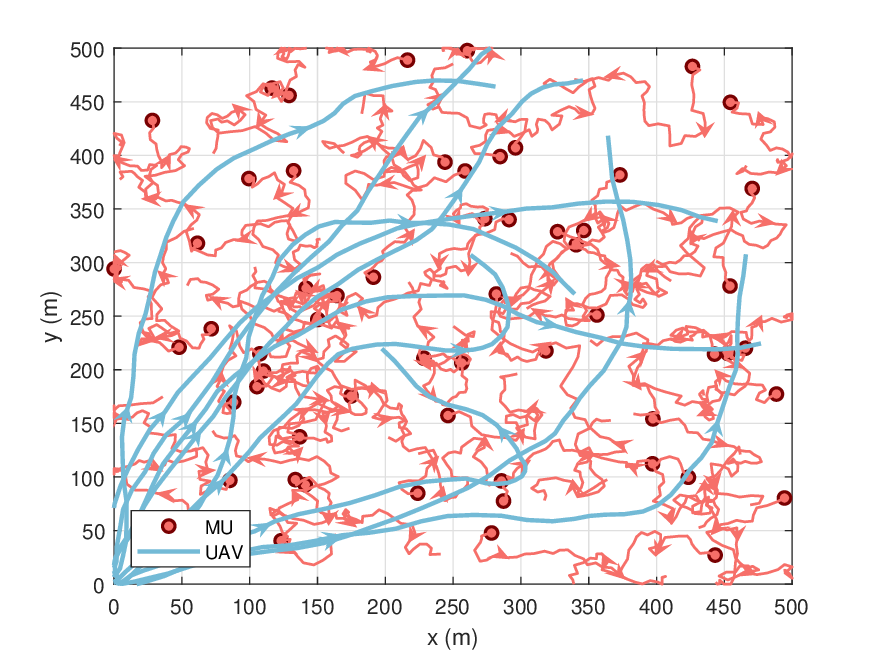}}
	\caption{The trajectories of UAVs for different scenarios.}
	\label{fig:trajectories}
\end{figure}

In Fig. \ref{fig:trajectories}, the trajectories of the MUs and the UAVs are showcased in different scenarios. Among which Fig. \ref{fig:trajectories}(a) presents a scenario where MUs are densely concentrated in a specific area of the region, while UAVs begin their operations from random positions. We observe that the trained UAVs adeptly navigate closer to MUs, hovering near the crowded areas to maximize the transmission rate. In Fig. \ref{fig:trajectories}(b), UAVs commence their operations from the lower left corner of the region, and MUs are relatively spread out. Initially, UAVs swiftly approach MUs and then collaborate to serve their respective associated MUs. Furthermore, UAVs prioritize closeness to their associated MUs as opposed to fully engaging with remote ones, demonstrating their ability to strategically consider long-term rewards. We illustrate the trajectory plots of UAVs under a large-scale MU scenario in Figs. \ref{fig:trajectories}(c) and (d). The UAVs navigate towards regions with a higher concentration of MUs, ensuring comprehensive coverage and equitable service provision. This further confirms that the reward effectively guides the UAVs to swiftly reach the MUs in larger-scale scenarios.

\section{Conclusion}\label{s:conclusion}
In the article, we studied the sensing, communication, and computation performance optimization in a multi-UAV-assisted ISCC network where data compression was exploited and MIMO radar sensing was designed. We proposed an MADRL-based energy consumption minimization to jointly optimize transmit beamforming, the compression and task partition factors, computational resource allocation, and UAV trajectory planning. To address the high-dimensional hybrid action spaces, the MAPPO method, which incorporated attention mechanism and Beta distribution, was used to attain the optimal strategy with effectiveness. The simulation results demonstrated that the proposed scheme can substantially reduce network energy consumption compared to the benchmark approaches. In future research, we will take into account the multi-device wireless interference framework for the inclusion of UAV relaying.

\bibliographystyle{IEEEtran}
\bibliography{IEEEabrv,refs}

\begin{thebibliography}{10}
\providecommand{\url}[1]{#1}
\csname url@samestyle\endcsname
\providecommand{\newblock}{\relax}
\providecommand{\bibinfo}[2]{#2}
\providecommand{\BIBentrySTDinterwordspacing}{\spaceskip=0pt\relax}
\providecommand{\BIBentryALTinterwordstretchfactor}{4}
\providecommand{\BIBentryALTinterwordspacing}{\spaceskip=\fontdimen2\font plus
\BIBentryALTinterwordstretchfactor\fontdimen3\font minus \fontdimen4\font\relax}
\providecommand{\BIBforeignlanguage}[2]{{%
\expandafter\ifx\csname l@#1\endcsname\relax
\typeout{** WARNING: IEEEtran.bst: No hyphenation pattern has been}%
\typeout{** loaded for the language `#1'. Using the pattern for}%
\typeout{** the default language instead.}%
\else
\language=\csname l@#1\endcsname
\fi
#2}}
\providecommand{\BIBdecl}{\relax}
\BIBdecl

\bibitem{9606720}
K.~B. Letaief, Y.~Shi, J.~Lu, and J.~Lu, ``Edge artificial intelligence for 6{G}: Vision, enabling technologies, and applications,'' \emph{IEEE Journal on Selected Areas in Communications}, vol.~40, no.~1, pp. 5--36, Jan. 2022.

\bibitem{9328305}
B.~Ji, Y.~Wang, K.~Song, C.~Li, H.~Wen, V.~G. Menon, and S.~Mumtaz, ``A survey of computational intelligence for 6{G}: Key technologies, applications and trends,'' \emph{IEEE Transactions on Industrial Informatics}, vol.~17, no.~10, pp. 7145--7154, Oct. 2021.

\bibitem{9282063}
H.~Yang, Z.~Wei, Z.~Feng, C.~Qiu, Z.~Fang, X.~Chen, and P.~Zhang, ``Queue-aware dynamic resource allocation for the joint communication-radar system,'' \emph{IEEE Transactions on Vehicular Technology}, vol.~70, no.~1, pp. 754--767, Jan. 2021.

\bibitem{9520318}
X.~Cao, B.~Yang, C.~Huang, C.~Yuen, Y.~Zhang, D.~Niyato, and Z.~Han, ``{Converged reconfigurable intelligent surface and mobile edge computing for space information networks},'' \emph{IEEE Network}, vol.~35, no.~4, pp. 42--48, Jul. 2021.

\bibitem{9737357}
F.~Liu, Y.~Cui, C.~Masouros, J.~Xu, T.~X. Han, Y.~C. Eldar, and S.~Buzzi, ``Integrated sensing and communications: Toward dual-functional wireless networks for 6{G} and beyond,'' \emph{IEEE Journal on Selected Areas in Communications}, vol.~40, no.~6, pp. 1728--1767, Jun. 2022.

\bibitem{9522072}
D.-H. Tran, V.-D. Nguyen, S.~Chatzinotas, T.~X. Vu, and B.~Ottersten, ``{UAV} relay-assisted emergency communications in {I}o{T} networks: Resource allocation and trajectory optimization,'' \emph{IEEE Transactions on Wireless Communications}, vol.~21, no.~3, pp. 1621--1637, Nov. 2022.

\bibitem{8579209}
B.~Li, Z.~Fei, and Y.~Zhang, ``{UAV} communications for 5{G} and beyond: Recent advances and future trends,'' \emph{IEEE Internet of Things Journal}, vol.~6, no.~2, pp. 2241--2263, Apr. 2019.

\bibitem{9943536}
C.~Wang, X.~Chen, J.~An, Z.~Xiong, C.~Xing, N.~Zhao, and D.~Niyato, ``Covert communication assisted by {UAV}-{IRS},'' \emph{IEEE Transactions on Communications}, vol.~71, no.~1, pp. 357--369, Nov. 2023.

\bibitem{9672750}
B.~Liu, Y.~Wan, F.~Zhou, Q.~Wu, and R.~Q. Hu, ``Resource allocation and trajectory design for {MISO} {UAV}-assisted {MEC} networks,'' \emph{IEEE Transactions on Vehicular Technology}, vol.~71, no.~5, pp. 4933--4948, May 2022.

\bibitem{data}
D.~Salomon, \emph{Data Compression: The Complete Reference.}\hskip 1em plus 0.5em minus 0.4em\relax New York, NY, {UAV}: Springer, 2004.

\bibitem{9648018}
Z.~Chen, N.~Zhao, D.~K.~C. So, J.~Tang, X.~Y. Zhang, and K.-K. Wong, ``Joint altitude and hybrid beamspace precoding optimization for {UAV}-enabled multiuser mmwave {MIMO} system,'' \emph{IEEE Transactions on Vehicular Technology}, vol.~71, no.~2, pp. 1713--1725, Dec. 2022.

\bibitem{8387798}
J.~Ren, G.~Yu, Y.~Cai, and Y.~He, ``Latency optimization for resource allocation in mobile-edge computation offloading,'' \emph{IEEE Transactions on Wireless Communications}, vol.~17, no.~8, pp. 5506--5519, Aug. 2018.

\bibitem{8635566}
D.~Xu, Q.~Li, and H.~Zhu, ``Energy-saving computation offloading by joint data compression and resource allocation for mobile-edge computing,'' \emph{IEEE Communications Letters}, vol.~23, no.~4, pp. 704--707, Apr. 2019.

\bibitem{9453853}
T.~Do-Duy, L.~D. Nguyen, T.~Q. Duong, S.~R. Khosravirad, and H.~Claussen, ``Joint optimisation of real-time deployment and resource allocation for {UAV}-aided disaster emergency communications,'' \emph{IEEE Journal on Selected Areas in Communications}, vol.~39, no.~11, pp. 3411--3424, Nov. 2021.

\bibitem{9622148}
B.~Dai, J.~Niu, T.~Ren, Z.~Hu, and M.~Atiquzzaman, ``Towards energy-efficient scheduling of {UAV} and base station hybrid enabled mobile edge computing,'' \emph{IEEE Transactions on Vehicular Technology}, vol.~71, no.~1, pp. 915--930, Jan. 2022.

\bibitem{9451579}
K.~Zhang, J.~Cao, and Y.~Zhang, ``Adaptive digital twin and multiagent deep reinforcement learning for vehicular edge computing and networks,'' \emph{IEEE Transactions on Industrial Informatics}, vol.~18, no.~2, pp. 1405--1413, Feb. 2022.

\bibitem{9828481}
Q.~Qi, X.~Chen, A.~Khalili, C.~Zhong, Z.~Zhang, and D.~W.~K. Ng, ``Integrating sensing, computing, and communication in 6{G} wireless networks: Design and optimization,'' \emph{IEEE Transactions on Communications}, vol.~70, no.~9, pp. 6212--6227, Sep. 2022.

\bibitem{9729765}
C.~Ding, J.-B. Wang, H.~Zhang, M.~Lin, and G.~Y. Li, ``Joint {MIMO} precoding and computation resource allocation for dual-function radar and communication systems with mobile edge computing,'' \emph{IEEE Journal on Selected Areas in Communications}, vol.~40, no.~7, pp. 2085--2102, Jul. 2022.

\bibitem{9916163}
Z.~Lyu, G.~Zhu, and J.~Xu, ``Joint maneuver and beamforming design for {UAV}-enabled integrated sensing and communication,'' \emph{IEEE Transactions on Wireless Communications}, vol.~22, no.~4, pp. 2424--2440, Apr. 2023.

\bibitem{9963915}
T.~Zhang, K.~Zhu, S.~Zheng, D.~Niyato, and N.~C. Luong, ``Trajectory design and power control for joint radar and communication enabled multi-{UAV} cooperative detection systems,'' \emph{IEEE Transactions on Communications}, vol.~71, no.~1, pp. 158--172, Nov. 2023.

\bibitem{9293257}
X.~Wang, Z.~Fei, J.~A. Zhang, J.~Huang, and J.~Yuan, ``Constrained utility maximization in dual-functional radar-communication multi-{UAV} networks,'' \emph{IEEE Transactions on Communications}, vol.~69, no.~4, pp. 2660--2672, Apr. 2021.

\bibitem{9739676}
K.~Meng, Q.~Wu, S.~Ma, W.~Chen, and T.~Q.~S. Quek, ``{UAV} trajectory and beamforming optimization for integrated periodic sensing and communication,'' \emph{IEEE Wireless Communications Letters}, vol.~11, no.~6, pp. 1211--1215, Jun. 2022.

\bibitem{9656117}
X.~Pang, N.~Zhao, J.~Tang, C.~Wu, D.~Niyato, and K.-K. Wong, ``{IRS}-assisted secure {UAV} transmission via joint trajectory and beamforming design,'' \emph{IEEE Transactions on Communications}, vol.~70, no.~2, pp. 1140--1152, Dec. 2022.

\bibitem{9729746}
B.~Chang, W.~Tang, X.~Yan, X.~Tong, and Z.~Chen, ``Integrated scheduling of sensing, communication, and control for {mmWave/THz} communications in cellular connected {UAV} networks,'' \emph{IEEE Journal on Selected Areas in Communications}, vol.~40, no.~7, pp. 2103--2113, Jul. 2022.

\bibitem{9847217}
S.~Hu, X.~Yuan, W.~Ni, and X.~Wang, ``Trajectory planning of cellular-connected {UAV} for communication-assisted radar sensing,'' \emph{IEEE Transactions on Communications}, vol.~70, no.~9, pp. 6385--6396, Aug. 2022.

\bibitem{10086052}
Y.~Qin, Z.~Zhang, X.~Li, W.~Huangfu, and H.~Zhang, ``Deep reinforcement learning based resource allocation and trajectory planning in integrated sensing and communications {UAV} network,'' \emph{IEEE Transactions on Wireless Communications}, vol.~22, no.~11, pp. 8158--8169, Nov. 2023.

\bibitem{9996408}
Z.~Wang, X.~Mu, Y.~Liu, X.~Xu, and P.~Zhang, ``{NOMA}-aided joint communication, sensing, and multi-tier computing systems,'' \emph{IEEE Journal on Selected Areas in Communications}, vol.~41, no.~3, pp. 574--588, Dec. 2023.

\bibitem{9424021}
L.~P. Qian, C.~Yang, H.~Han, Y.~Wu, and L.~Meng, ``Learning driven resource allocation and {SIC} ordering in {EH} relay aided {NB}-{IoT} networks,'' \emph{IEEE Communications Letters}, vol.~25, no.~8, pp. 2619--2623, Aug. 2021.

\bibitem{9840900}
N.~Huang, T.~Wang, Y.~Wu, Q.~Wu, and T.~Q.~S. Quek, ``Integrated sensing and communication assisted mobile edge computing: An energy-efficient design via intelligent reflecting surface,'' \emph{IEEE Wireless Communications Letters}, vol.~11, no.~10, pp. 2085--2089, Oct. 2022.

\bibitem{8663615}
Y.~Zeng, J.~Xu, and R.~Zhang, ``Energy minimization for wireless communication with rotary-wing {UAV},'' \emph{IEEE Transactions on Wireless Communications}, vol.~18, no.~4, pp. 2329--2345, 2019.

\bibitem{9254093}
H.~Peng and X.~Shen, ``Multi-agent reinforcement learning based resource management in {MEC}- and {UAV}-assisted vehicular networks,'' \emph{IEEE Journal on Selected Areas in Communications}, vol.~39, no.~1, pp. 131--141, Jan. 2021.

\bibitem{9404260}
C.~Zhan, H.~Hu, Z.~Liu, Z.~Wang, and S.~Mao, ``Multi-{UAV}-enabled mobile-edge computing for time-constrained {IoT} applications,'' \emph{IEEE Internet of Things Journal}, vol.~8, no.~20, pp. 15\,553--15\,567, Oct. 2021.

\bibitem{7542156}
Y.~Wang, M.~Sheng, X.~Wang, L.~Wang, and J.~Li, ``Mobile-edge computing: Partial computation offloading using dynamic voltage scaling,'' \emph{IEEE Transactions on Communications}, vol.~64, no.~10, pp. 4268--4282, Oct. 2016.

\bibitem{Chou2017ImprovingSP}
P.-W. Chou, D.~Maturana, and S.~A. Scherer, ``Improving stochastic policy gradients in continuous control with deep reinforcement learning using the beta distribution,'' in \emph{{Proc.} ICML}, 2017, pp. 834--843.

\bibitem{9583941}
T.~Cai, Z.~Yang, Y.~Chen, W.~Chen, Z.~Zheng, Y.~Yu, and H.-N. Dai, ``Cooperative data sensing and computation offloading in {UAV}-assisted crowdsensing with multi-agent deep reinforcement learning,'' \emph{IEEE Transactions on Network Science and Engineering}, vol.~9, no.~5, pp. 3197--3211, Oct. 2022.

\bibitem{9791349}
C.~Liu, W.~Yuan, S.~Li, X.~Liu, H.~Li, D.~W.~K. Ng, and Y.~Li, ``Learning-based predictive beamforming for integrated sensing and communication in vehicular networks,'' \emph{IEEE Journal on Selected Areas in Communications}, vol.~40, no.~8, pp. 2317--2334, Jun. 2022.

\bibitem{10158711}
Z.~He, W.~Xu, H.~Shen, D.~W.~K. Ng, Y.~C. Eldar, and X.~You, ``Full-duplex communication for {ISAC}: Joint beamforming and power optimization,'' \emph{IEEE Journal on Selected Areas in Communications}, vol.~41, no.~9, pp. 2920--2936, Jun. 2023.

\bibitem{9209079}
L.~Wang, K.~Wang, C.~Pan, W.~Xu, N.~Aslam, and L.~Hanzo, ``Multi-agent deep reinforcement learning-based trajectory planning for multi-{UAV} assisted mobile edge computing,'' \emph{IEEE Transactions on Cognitive Communications and Networking}, vol.~7, no.~1, pp. 73--84, Mar. 2021.

\bibitem{8807184}
H.~Sun, W.~Shi, X.~Liang, and Y.~Yu, ``Vu: Edge computing-enabled video usefulness detection and its application in large-scale video surveillance systems,'' \emph{IEEE Internet of Things Journal}, vol.~7, no.~2, pp. 800--817, Aug. 2020.

\bibitem{10021296}
W.~Liu, B.~Li, W.~Xie, Y.~Dai, and Z.~Fei, ``Energy efficient computation offloading in aerial edge networks with multi-agent cooperation,'' \emph{IEEE Transactions on Wireless Communications}, vol.~22, no.~9, pp. 5725--5739, Sep. 2023.

\bibitem{8956055}
Z.~Yu, Y.~Gong, S.~Gong, and Y.~Guo, ``Joint task offloading and resource allocation in {UAV}-enabled mobile edge computing,'' \emph{IEEE Internet of Things Journal}, vol.~7, no.~4, pp. 3147--3159, Jan. 2020.

\end{thebibliography}

\end{document}